\documentclass[sigconf,10pt]{acmart} %
\usepackage{xcolor}
\usepackage{hyperref}
\usepackage{hyperxmp}
\usepackage{xurl}
\usepackage{booktabs}
\usepackage[ruled]{algorithm2e}
\usepackage{enumitem}
\usepackage{enumerate}
\setlist{noitemsep,topsep=0pt,parsep=0pt,partopsep=0pt}
\usepackage[english]{babel}
\usepackage{marvosym}
\usepackage[font=small,skip=5pt]{caption}
\usepackage{amsmath}

\usepackage{amssymb}
\usepackage{xspace}
\usepackage{multirow}
\usepackage{romannum}
\usepackage{threeparttable}
\usepackage{float}
\usepackage{flafter}
\usepackage{comment}
\usepackage{graphicx}
\usepackage{pifont}
\usepackage{algpseudocode}
\usepackage{arydshln}
\usepackage{lineno}

\def\fig{Fig.\xspace}
\def\eqn{Eqn.\xspace}

\def\tab{Tab.\xspace}

\def\ie{{\textit{i.e.}\xspace}} 
\def\eg{{\textit{e.g.}\xspace}}

\def\vs{{\textit{vs.}\xspace}}

\newcommand{\head}[1]{{\noindent \textbf{#1:}}}

\graphicspath{{./pdf/}}
\DeclareGraphicsExtensions{.pdf,.jpeg,.png}

\ifodd 0
\newcommand{\rev}[1]{#1}%
\newcommand{\com}[1]{\textbf{\color{red}(COMMENT: #1)}} %
\newcommand{\todo}[1]{\textbf{{\color{orange}(TODO: #1)}}}
\else

\newcommand{\rev}[1]{#1}
\newcommand{\com}[1]{}
\newcommand{\todo}[1]{}
\fi

\newcommand{\bs}[1]{\boldsymbol{#1}}

\newcommand{\mc}[1]{\mathcal{#1}}
\newcommand{\mf}[1]{\mathfrak{#1}}
\newcommand{\trans}{\mathsf{T}}
\newcommand{\ctrans}{\mathsf{H}}
\DeclareMathOperator*{\argmin}{arg\,min}

\acmYear{2026}\copyrightyear{2026}
\setcopyright{cc}
\setcctype[4.0]{by}
\acmConference[MobiCom '26]{The 32nd Annual International Conference on Mobile Computing and Networking}{October 26--30, 2026}{Austin, TX, USA}
\acmBooktitle{The 32nd Annual International Conference on Mobile Computing and Networking (MobiCom '26), October 26--30, 2026, Austin, TX, USA}
\acmDOI{10.1145/3795866.3796683}
\acmISBN{979-8-4007-2505-0/26/10}

\begin{document}

\def\sysname{\textsf{RF-LEGO}\xspace}
\newcommand{\blueurl}[1]{
    \href{#1}{\textcolor{blue}{\texttt{#1}}}\xspace
}
\def\codeurl{\blueurl{https://github.com/aiot-lab/RF-LEGO}}

\title[\sysname]{\sysname: Modularized Signal Processing-Deep Learning Co-Design for RF Sensing via Deep Unrolling}

\author{Luca Jiang-Tao Yu}
\affiliation{ 
    \institution{The University of Hong Kong}
    \city{}\country{}
}
\email{lucayu@connect.hku.hk}

\author{Chenshu Wu}
\affiliation{ 
    \institution{The University of Hong Kong}
    \city{}\country{}
}
\email{chenshu@cs.hku.hk}

\renewcommand{\shortauthors}{Yu and Wu}
\renewcommand{\authors}{Luca Jiang-Tao Yu, Chenshu Wu}

\begin{abstract}
Wireless sensing, traditionally relying on signal processing (SP) techniques, has recently shifted toward data-driven deep learning (DL) to achieve performance breakthroughs. 
However, existing deep wireless sensing models are typically end-to-end and task-specific, lacking \rev{reusability and interpretability}. 
We propose \sysname, a modular co-design framework that transforms interpretable SP algorithms into trainable, physics-grounded DL modules through deep unrolling. 
By replacing hand-tuned parameters with learnable ones while preserving core processing structures and mathematical operators, \sysname ensures modularity, cascadability, and structure-aligned interpretability. 
Specifically, we introduce three deep-unrolled modules for critical RF sensing tasks: frequency transform, spatial angle estimation, and signal detection. 
Extensive experiments \rev{using real-world data for Wi-Fi, millimeter-wave, UWB, and 6G sensing} 
demonstrate that \sysname significantly outperforms existing SP and DL baselines, both standalone and when integrated into multiple downstream tasks. 
\sysname pioneers a novel SP-DL co-design paradigm for wireless sensing via deep unrolling, shedding light on efficient and interpretable deep wireless sensing solutions.
Our code is available at \codeurl.
\end{abstract}

\begin{CCSXML}
<ccs2012>
   <concept>
       <concept_id>10003120.10003138</concept_id>
       <concept_desc>Human-centered computing~Ubiquitous and mobile computing</concept_desc>
       <concept_significance>500</concept_significance>
       </concept>
   <concept>
       <concept_id>10003033.10003106.10003113</concept_id>
       <concept_desc>Networks~Mobile networks</concept_desc>
       <concept_significance>500</concept_significance>
       </concept>
 </ccs2012>
\end{CCSXML}

\ccsdesc[500]{Human-centered computing~Ubiquitous and mobile computing}
\ccsdesc[500]{Networks~Mobile networks}

\keywords{Deep Unrolling, Wireless Sensing, Signal Processing, Deep Learning, RF Signals}

\maketitle
\begin{figure}[t!]
    \centering
    \includegraphics[width=0.9\linewidth]{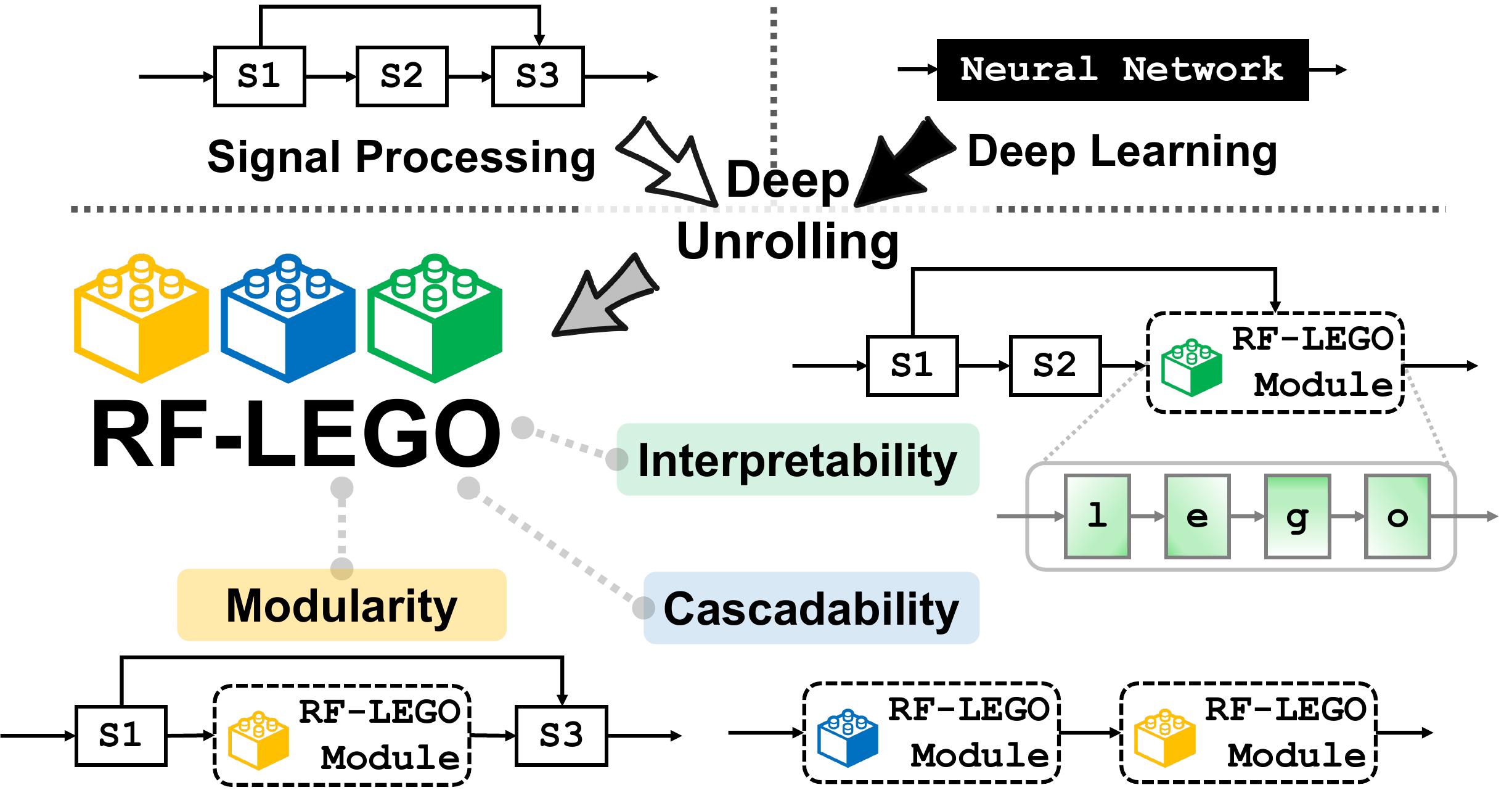}
    \caption{Core principles of \sysname: Modularity, Cascadability, and Interpretability. \rm{\sysname bridges the gap between classical SP and DL for RF sensing via deep unrolling.}}
    \label{fig:fig1}
\end{figure}

\section{Introduction}
\label{sec:introduction}

Over the past decade, wireless sensing has transformed the wireless industry across Wi-Fi, millimeter-wave/UWB radars, 6G networks, LoRa, and even GPS. 
As AI continues to revolutionize various applications, the field of wireless sensing is undergoing a significant shift from traditional model-centric signal processing \cite{zhang2019smars,lyu2024ase, zhang2023vecare} toward data-heavy deep learning \cite{zhao2023nerf2,zhao2023radio2text,LiquImager,yu2025uspeech}. 
Early efforts in Deep Wireless Sensing (DWS) \cite{ozturk2023radio, zhao2025space} exploit purely data-driven approaches by leveraging existing neural network architectures like CNNs, RNNs, and Transformers, with recent attempts to explore large language models (LLMs) to process wireless signals directly \cite{ji2024hargpt,zhang2025wi,zhang2025sensorlm}. 
Albeit inspiring, these models face significant challenges related to data scarcity and model interpretability, which in turn hinder generalization, reuse, and deployment efficiency.
To overcome these issues, researchers have resorted to signal-informed models through signal processing-deep learning (SP-DL) co-design, and remarkable advances have been achieved: 
(i) direct embedding of signal processing blocks to inject physical priors into neural networks \cite{ding2020rf}; (ii) RF-intrinsic basis-inspired design, \eg, phase-aware encoder \cite{chi2024rf,yang2023slnet}; (iii) DSP-inspired parameterizations and mechanisms, like Fourier-based initialization \cite{zhao2023cubelearn} and STFT-like operations \cite{li2021units, yao2019stfnets}; and (iv) RF-specific architectures with math-interpretable design \cite{zhang2025unlocking,diskin2024cfarnet}. 
Despite the progress, existing designs still suffer from major limitations: 
\textit{1) Limited Reusability}: Most models are trained end-to-end for a specific downstream task, such as gesture recognition or vital sign monitoring \cite{li2021units, lin2019dl, ding2020rf, zhao2023cubelearn, chen2021movi,zheng2021more}. This undermines their reusability, making it difficult to reuse or swap individual modules when tasks or environments vary. Unlike in computer vision or natural language processing, few DWS models are reused in subsequent research. 
\textit{2) Lack of Interpretability}: 
Although some prior works incorporate signal processing inspirations in network design \cite{lin2019dl,yang2023slnet}, they lack structure-aligned interpretability; specifically, they do not maintain a clear stage-by-stage signal processing diagram with well-defined input/output contracts and semantically meaningful intermediate outputs in standard signal domains.
Recently, \textit{deep unrolling} techniques have emerged as a promising approach to SP-DL co-design by unrolling classical algorithms in a learnable framework. 
Deep unrolling replaces fixed algorithmic coefficients with learnable yet physics-constrained parameters while preserving operator structure and physical input semantics \cite{monga2021algorithm}, reserving the operator interpretability of signal processing while leveraging the parameter learnability of deep learning. 
These compelling advantages inspire us to embrace deep unrolling for \textit{modularized SP-DL co-design}, promoting both reusability and interpretability for DWS models. 
While deep unrolling has been successful in numerous classical algorithms \cite{revach2022kalmannet,merkofer2023music,lin2019dl,liang2025cfarnet,li2020efficient,singhal1988training}, modularized unrolling in DWS faces unique challenges, as summarized below: 
Different signal processing operators impose distinct proximal structures and constraints; non-differentiable steps require smooth, bounded surrogates to prevent back-propagation collapse; and true modularity demands preserving the classical input-output and complex-valued semantics so blocks remain plug-and-play and calibratable across platforms.

We propose \sysname, as shown in \fig \ref{fig:fig1}, the first modularized SP-DL co-design approach for RF sensing based on deep unrolling. By turning each algorithmic iteration into a differentiable, trainable layer, \sysname maintains mathematically grounded SP interfaces while replacing handcrafted configurations with learnable parameters, delivering reusable SP-DL modules with structure-aligned interpretability. 
The design of \sysname centers around three key principles: 

\head{$\bullet$ Modularity} Our modules unroll classical algorithms, mirroring their input/output semantics while introducing a minimal number of trainable parameters. By doing so, each module will be plug-and-play across different pipelines, datasets, and applications without requiring retraining. 

\head{$\bullet$ Cascadability} The modular nature of these blocks allows them to be flexibly combined with each other or integrated into existing methods, offering reusable \emph{"LEGO bricks"} for effortlessly building sophisticated learning pipelines. 

\head{$\bullet$ Interpretability} By converting algorithmic iterations into differentiable and trainable layers, the proposed blocks preserve SP-aligned operator structures and semantically meaningful intermediate outputs. 
While this design does not ensure strict model interpretability of the neural networks and the learned parameters, it offers intra-block structure and inter-block pipeline interpretability as they are precisely aligned with classical SP pipelines.

Among various signal processing techniques involved in RF sensing, we primarily focus on the three most fundamental components, \ie, frequency transformation, spatial angle estimation, and signal detection, which have been widely used in most of the existing literature \cite{ozturk2023radio,zhao2025space,zhao2023radio2text,wang2020vimo,zhang2019smars,zheng2021more,dodds2025non,zhao2018rf,zheng2019zero,yang2023slnet,yao2019stfnets,zhao2023cubelearn,lai2024enabling,LiquImager,freegait-tmc}. 
Correspondingly, we present the following three deep unrolled algorithms in \sysname, which can already enable many RF sensing applications. We leave it to future work in the community to contribute more \sysname blocks for other useful techniques.

\head{$\blacksquare$ \sysname FT}
Frequency Transformation (FT) is a fundamental processing step in almost all RF sensing applications.
However, the classical Cooley-Tukey FFT \cite{cooley1965algorithm} has fixed basis functions, rendering it inherently rigid and unable to adapt to signal-dependent artifacts like spectral leakage, which can obscure weak targets in noisy environments.
To overcome this, we choose to deep unroll \emph{Bluestein's Algorithm} implementation \cite{bluestein2003linear}, which factorizes the transform into chirp multiplications and a single convolution.
This exposes a clean insertion point for a compact, complex-valued learnable operator, allowing for data-driven suppression of spectral leakage while preserving the intrinsic semantics of RF signals.

\head{$\blacksquare$ \sysname Beamformer} 
RF sensing often performs a spatial transformation to obtain the angle spectrum, a crucial step to separate multiple targets. 
Super-resolution methods like MUSIC \cite{schmidt1986multiple}, relying on subspace decomposition, are notoriously brittle in practice, failing under challenging conditions like coherent multipath and low SNR conditions. 
Furthermore, it faces unstable gradients during backpropagation, often causing training to fail if unrolled. 
In \sysname, we instead investigate the unrolled LASSO beamformer \cite{malioutov2005sparse} using ADMM (Alternating Direction Method of Multipliers) \cite{boyd2011distributed}. This iterative framework is inherently better suited for real-world conditions. 
While the ADMM solver using rigid parameters and fixed update rules exhibits limited performance and slow convergence in complex scenarios \cite{wu2020sparse}, our unrolled design addresses this by making these coefficients trainable while leaving the essential steering model unchanged.

\head{$\blacksquare$ \sysname Detector}
Many classical approaches, like the commonly used Constant False Alarm Rate (CFAR) family \cite{Hansen1973,weiss2007analysis,truck1977range,rohling2007radar}, have been developed for signal detection, an essential step for sensing applications. 
The core operations of CFAR detectors, however, are non-differentiable and hence cannot be directly unrolled. 
For example, operators like order statistics, used for robust noise estimation, are incompatible with gradient-based backpropagation. 
We therefore propose an unrolled design with the same workflow as classical CFAR, yet recast its adaptive noise estimation as a compact and differentiable state space model \cite{hamilton1994state}, which acts as a smooth surrogate and allows the detector to learn the adaptive dynamics. 
The design not just circumvents the tedious manual parameter configurations in classical CFAR, but also improves robustness across diverse environments.
There are existing works to improve these classical algorithms with deep learning, which, however, are mostly end-to-end models lacking interpretability \cite{zhao2023cubelearn,merkofer2023music,lin2019dl}. 
In contrast, we propose novel techniques to unroll classical algorithms as learnable blocks without changing the original signal processing structure. 
Note that our designs are primarily within the context of wireless sensing, rather than for generic applications like the original algorithms.

We conduct extensive real-world experiments to evaluate \sysname, with a main focus on validating the effectiveness of individual unrolled techniques, the performance of applying \sysname techniques in tandem with other existing models, and the performance of downstream tasks such as trajectory tracking, vital sign monitoring, and human activity recognition.
We evaluate on both self-collected data and public datasets, totaling approximately 3M frames across various environments and different modalities (including Wi-Fi, millimeter-wave, UWB, and 6G). 
Our experimental results demonstrate that \sysname consistently outperforms classical SP baselines and achieves competitive performance compared with prior DL baselines, while offering reusable learnable components for broader applications. 
\sysname pioneers a new paradigm for building reusable and interpretable SP-DL co-design for wireless sensing via deep unrolling, inspiring new research opportunities in the community to build generalizable and efficient DWS solutions.

In summary, our main contributions are as follows:
\begin{itemize}[noitemsep, topsep=0pt, leftmargin=*]
    \label{contribution}
    \item We present \sysname, the first modular co-design framework that bridges rigid-but-interpretable signal processing and adaptive-but-opaque deep learning via \emph{deep unrolling}, yielding reusable and interpretable learning modules for the wireless sensing community.
    
    \item We propose three deep unrolled, physics-grounded modules, \ie, \sysname FT for robust frequency transformation, \sysname Beamformer for angle estimation, and \sysname Detector for adaptive signal detection.
    
    \item We conduct extensive evaluation using real-world mmWave, UWB, and Wi-Fi signals, demonstrating the effectiveness of \sysname techniques either when using individually or integrating into existing networks. 
\end{itemize}

\begin{figure}[t]
    \centering
    \includegraphics[width=1\linewidth]{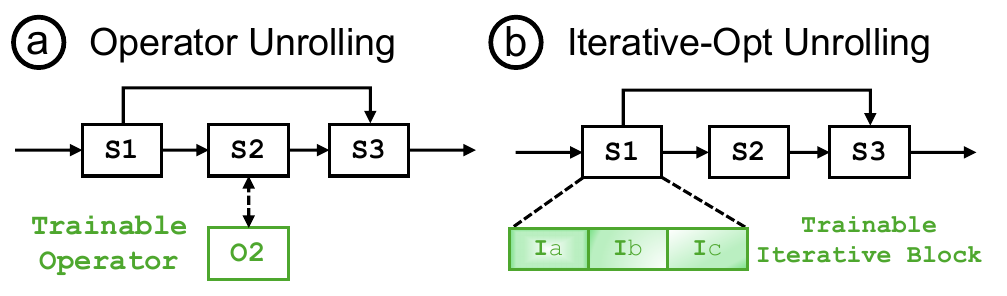}
    \caption{Deep Unrolling. \rm{\texttt{S}: Signal Processing Block; \texttt{O}: Unrolled Trainable Operator; \texttt{I}: Unrolled Trainable Iterative Block.}}
    \label{fig:deep_unrolling}
\end{figure}

\section{A Primer on Deep Unrolling}
\label{sec:deep_unrolling}

Most deep learning models are purely data-driven, and their learned structures are difficult to interpret. End-to-end networks learn task-specific mappings (\eg, regression or classification) entirely through backpropagation over the high-dimensional parameters whose individual roles are opaque \cite{monga2021algorithm}. By contrast, classical signal processing is interpretable at both the pipeline and block levels because it is derived from physical models and domain priors, \ie, the intrinsics of signals. However, traditional methods suffer from parameter rigidity, meaning that their performance hinges on expert-tuned hyperparameters that often need to be recalibrated upon environment changes.
Deep unrolling transforms signal processing techniques into structured neural architectures, combining the adaptability of deep learning with the interpretability and physics of classical methods. Broadly, as illustrated in \fig \ref{fig:deep_unrolling}, it falls into two categories as follows:

\begin{itemize}[noitemsep, topsep=0pt, leftmargin=*]

    \item \textbf{Operator Unrolling}.  Keep the signal processing pipeline, but replace a classical block with a drop-in trainable operator that mirrors its mathematics. Fixed coefficients (\eg, kernels, filters, thresholds) become a compact set of physics-constrained learnable parameters, while complex-valued semantics and intermediate states remain exposed. The result is a plug-and-play block with strong domain alignment, interpretability, and easy composition with neighboring stages.

    \item \textbf{Iterative-Optimization Unrolling}. Keep the classical block but open its inner loop: each solver's iterative block becomes a network layer with learnable step sizes, preconditioners, and proximal or shrinkage maps. This turns an algorithm into a structured, trainable stack that preserves the structure-alignedinterpretability.
\end{itemize}

Together, these two categories provide a concise design vocabulary: either swap in a trainable operator that respects the classical interface, or unfold the operator’s iterative routine into a stable, interpretable, and data-driven network. Subsequent sections instantiate these patterns for specific RF sensing blocks. Deep unrolling has achieved remarkable advances in various domains, with successful applications in unrolling classical algorithms for sparse coding \cite{wu2020sparse}, Kalman filtering \cite{revach2022kalmannet, singhal1988training}, and image restoration \cite{li2020efficient}. However, its potential has not yet been fully explored for the unique challenges of DWS. 
\sysname pioneers this direction by proposing a framework for deep unrolled RF sensing.

\section{\sysname Design}
\label{sec:design}

\subsection{\sysname FT}
\label{sec:sysname_frequency_transformation}

Time-frequency transformation is a cornerstone of wireless sensing for extracting range and Doppler information.
Classical methods such as FFT rely on fixed bases, making the spectrum vulnerable to significant sidelobe leakage. In challenging scenarios with multipath fading, this artifact can obscure weaker signals or even submerge the main lobe. 
Their coefficients are scene independent, so robustness and adaptivity degrade across hardware and environments in RF sensing. 
The widely-used Cooley-Tukey implementation is highly optimized for the common $2^N$ case, but its multi-stage factorization offers no localized, structure-preserving learning handle.
Alternatively, Bluestein's Algorithm recasts the transformation as a convolution with a fixed chirp, which is efficient but not adaptive to signal-dependent artifacts.
Preprocessing with filters and heuristic parameter adjustments provides limited gains, while advanced decompositions such as the Hilbert-Huang transform \cite{huang1998empirical} increase computational cost and are susceptible to mode mixing.
Meanwhile, many end-to-end models simply adopt vanilla architectures that ignore both the temporal-frequency domain gap and the complex-valued nature of RF signals \cite{zhao2023cubelearn}.
Below, we outline our proposed design to address these limitations. 

\begin{figure}[t]
    \centering
    \includegraphics[width=\linewidth]{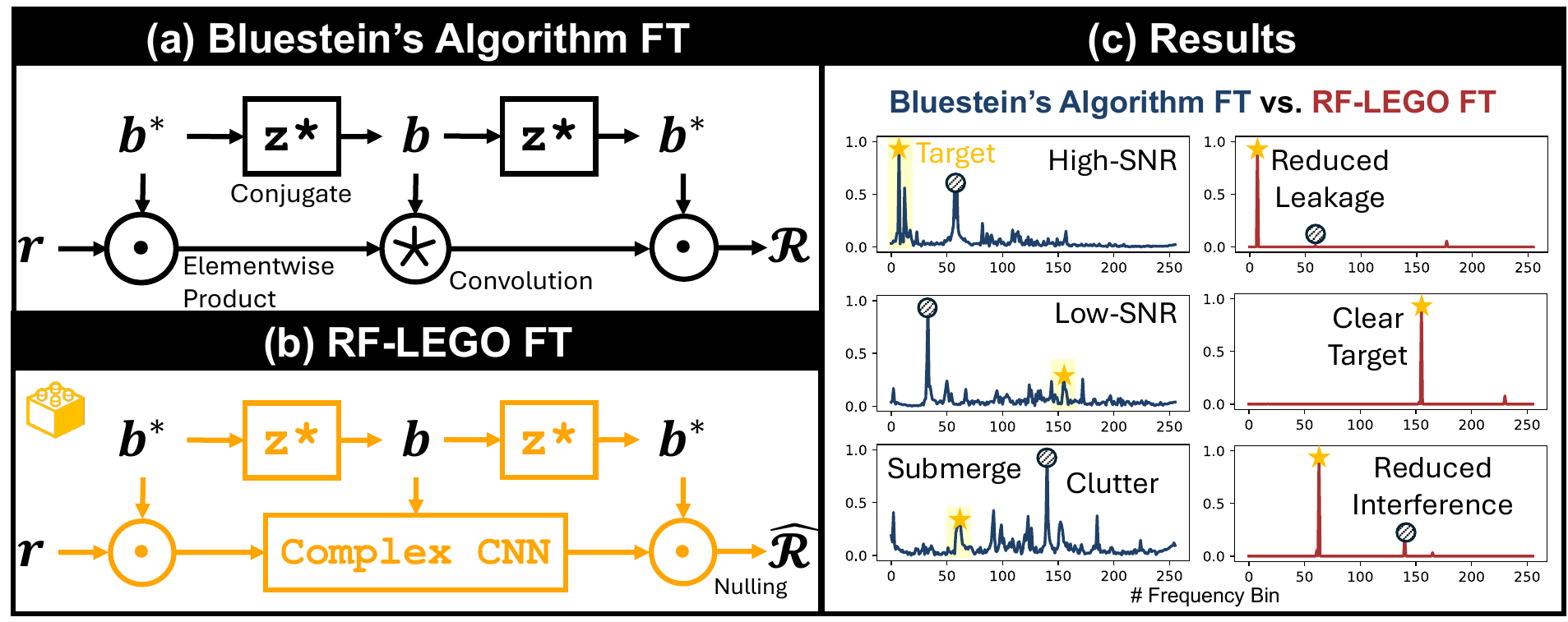}
    \caption{\sysname FT. \rm{(a) Bluestein's Algorithm implementation of Fourier transformation via convolution, where $\bs{b}$ represents the chirp signal and $\bs{z^*}$ denotes its conjugate. (b) \sysname FT replaces the fixed convolution with a learnable convolutional layer.}}
    \label{fig:lego_ft}
\end{figure}

\noindent \textbf{Signal Model}.
Let $\bs{s}\in\mathbb{C}^{N}$ denote a clean complex-baseband signal with samples $\{s_n\}_{n=0}^{N-1}$, and let $\bs{n}\in\mathbb{C}^{N}$ be additive noise. 
The received signal is $\bs{r}=\bs{s}+\bs{n}$. 
Let $\mf{F}(\cdot)$ be the discrete Fourier Transformation. 
The frequency-domain samples $\{\mathcal{R}[k]\}_{k=0}^{N-1}$ satisfy
\begin{equation}
    \begin{aligned}
        \mc{R}[k] = \mf{F}(\bs{s}+\bs{n})[k] = \sum_{n=0}^{N-1} \bs{s}_n e^{-j 2 \pi k \frac{n}{N}} + \mf{F}(\bs{n})[k],
    \end{aligned}
    \label{eqn:fft}
\end{equation}
where $k=0, \dots, N-1,$ and the noise term perturbs both amplitude and phase. We do not impose a specific distribution on $\bs{n}$ and treat $\mf{F}(\bs{n})$ through its statistics.
To expose a structure that preserves the classical transformation while enabling learning, we adopt Bluestein's Algorithm implementation, which is rewritten as a convolution with a known chirp. Using the identity: $n \cdot k=-\tfrac{(k-n)^2}{2}+\tfrac{n^2}{2}+\tfrac{k^2}{2}$, we obtain
\begin{equation}
    \begin{aligned}
        \mc{R}[k] 
        &= e^{-j \pi \frac{k^2}{N}} \cdot (\bs{s}_ne^{-j \pi \frac{n^2}{N}} \circledast e^{j \pi \frac{n^2}{N}})[k] + \mf{F}(\bs{n})[k],
    \end{aligned}
    \label{eqn:ba_fft}
\end{equation}
with $\bs{a}_n=\bs{s}_n e^{-j\pi n^2/N}$, $\bs{b}_n=e^{j\pi n^2/N}$, and $\circledast$ denoting discrete convolution. 
This factorization is algebraically equivalent to the discrete Fourier Transformation, retains the complex-valued semantics, and isolates a single fixed-kernel convolution as illustrated in \fig \ref{fig:lego_ft}(a). 

\noindent\textbf{Deep Unrolling.}
The convolutional form in \eqn \eqref{eqn:ba_fft} reveals a practical handle: the discrete Fourier Transformation can be executed as a single convolution with a known chirp.
In real scenarios, however, clutter and other non-stationary, signal-dependent disturbances limit what a fixed chirp can achieve. We therefore keep the outer chirp multiplications intact and make only the convolution learnable, yielding a frequency-transformation block that remains structurally identical to the classical operator and preserves complex-valued architecture.
Concretely, deep learning toolchains implement cross-correlation rather than flipped-kernel convolution as follows: 
\begin{equation}
\label{eqn:comparison_sp_dl}
    (\bs{a} \circledast \bs{b})[k] \triangleq
    \begin{cases}
        \sum_{n=0}^{N-1}\bs{a}[k-n] \cdot \bs{b}[n], & \text{(SP)} \\
        \sum_{n=0}^{N-1}\bs{a}[k+n] \cdot \bs{b}[n], & \text{(DL)}\\
    \end{cases}
    .
\end{equation}
We exploit the mathematical equivalence: a correlation with a filter is identical to a convolution with the flipped filter if trainable \cite{bellanger2000digital}. 
Thus, we replace the signal processing convolution by a hierarchical learnable complex filter applied through correlation, with weights initialized from the chirp sequence $\bs{b}$. 
This preserves the algebraic blueprint of Bluestein's Algorithm while exposing a compact set of trainable coefficients.
We instantiate the learnable operator as a shallow complex-valued convolutional neural network that sits exactly at the convolution site of Bluestein’s Algorithm FT. 
The network uses smooth activations to provide mild nonlinearity and is initialized from signal processing bases to ensure a stable start and domain consistency. 
An optional, lightweight nulling head applies a soft, trainable shrinkage at the output to attenuate residual leakage. 
Our ablation study (\S\ref{sec:ablation_study}) shows that the main gains stem from the unrolled convolution itself, with the nulling head acting as a stabilizer rather than the primary source of improvement.
This design keeps interpretability, retains complex-valued semantics, and introduces data-driven adaptation where classical methods are rigid. In practice, it improves spectral suppression and reduces leakage while remaining plug-and-play with other modules. As visualized in \fig \ref{fig:lego_ft} on the mmWave range frequency axis, \sysname FT effectively suppresses sidelobes in low-SNR conditions and isolates targets submerged in clutter where the classical algorithm fails.

\subsection{\sysname Beamformer}
\label{sec:sysname_beamforming}
\begin{figure}[t]
    \centering
    \includegraphics[width=\linewidth]{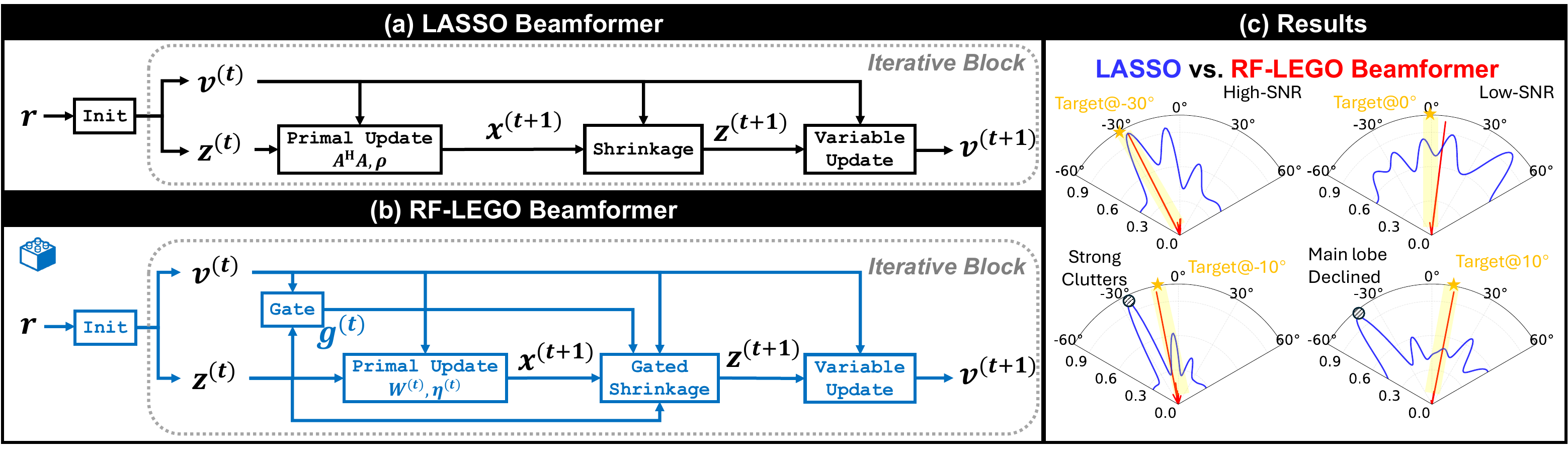}
    \caption{\sysname Beamformer. \rm{(a) The classical LASSO solver using ADMM. (b) The unrolled architecture with learnable parameters. Here, $\bs{x^{(t)}}$ represents the estimated sparse angular spectrum, $\bs{z^{(t)}}$ is the auxiliary variable, $\bs{v^{(t)}}$ is the dual variable, and $\bs{g^{(t)}}$ denotes the learned gate that dynamically balances the update between historical and current estimates.}}
    \label{fig:lego_beamformer}
\end{figure}

Beamforming converts array measurements into an angular spectrum for estimating the angle of arrival. 
Many signal processing techniques have been proposed, including the classical delay and sum (DAS) and Bartlett \cite{trees2002optimum}, Capon \cite{capon1969high}
beamformers, and subspace methods such as Multiple Signal Classification (MUSIC) \cite{schmidt1986multiple}. 
Built upon certain signal models, these methods suffer from different limitations. 
For example, the most widely used MUSIC method requires the number of sources and assumes a clear separation between the signal and noise subspaces. 
Moreover, in practice, these methods rely on strong modeling assumptions and manual tuning, which limits robustness across hardware and environments. 
Recent attempts exploit deep neural networks for better performance. Yet existing works are mostly end-to-end models for specific tasks. They often do not provide an explicit angle spectrum \cite{zhao2023cubelearn}.
Furthermore, some models incorporate operations that are ill-suited for gradient-based training; for instance, methods based on the MUSIC algorithm require eigenvalue decomposition, which is known to have unstable gradients that can destabilize the training process \cite{merkofer2023music,pytorch2024eig}.

These drawbacks motivate a tightly coupled unrolled design that keeps the classical steering model and explicit angle spectrum, avoids unstable subspace factorization, and learns only a compact set of physics-constrained solver coefficients that are otherwise hand-tuned.
We recast beamforming as sparse spectral estimation on a discretized set of angles and adopt the Least Absolute Shrinkage and Selection Operator (LASSO) formulation \cite{malioutov2005sparse} to enforce a sparse angle spectrum consistent with the steering model.
Based on this formulation, we then employ iterative optimization unrolling to unroll the LASSO beamformer as an improved Alternating Direction Method of Multipliers (ADMM) solver \cite{boyd2011distributed}. 
The unrolled structure keeps the classical operator and complex-valued inputs intact while converting a compact set of physics-constrained coefficients, such as step sizes, preconditioning weights, and shrinkage levels, into learnable parameters. 
This design outperforms conventional methods by removing reliance on subspace factorization and yielding robust performance under low SNR and coherent sources, while retaining an interpretable structure and a reusable interface like the classical methods.
Our detailed signal model and unrolled design are as follows.

\noindent \textbf{Signal Model}. 
Consider a uniform linear array (ULA) with $M$ elements spaced at $\lambda/2$. The array observes $K$ far-field sources from directions $\{\bs{\theta}_0,\bs{\theta}_1,\ldots,\bs{\theta}_{K-1}\}$ in spatially white complex Gaussian noise $\bs{n}\sim\mathcal{CN}(0,\sigma^2\bs{I})$. The received snapshot $\bs{r}\in\mathbb{C}^{M}$ is modeled as
\begin{equation}
    \bs{r} = \sum_{k=0}^{K-1} \bs{s}[k] \bs{a}(\bs{\theta}_k) + \bs{n},
\end{equation}
where $\bs{s}\in\mathbb{C}^{K}$ collects the complex source responses and $\bs{a}(\bs{\theta}_k)\in\mathbb{C}^{M}$ is the steering vector $\bs{a}(\bs{\theta}_k)^{\trans} = [e^{-j\frac{2\pi \bs{d}}{\lambda}i \sin{\bs{\theta}_k}}]^{M-1}_{i=0}$.
Our goal is to estimate a sparse angle spectrum $\hat{\bs{x}}\in\mathbb{C}^{N}$ on a predefined grid $\{\bs{\theta}_0,\ldots,\bs{\theta}_{N-1}\}$ by solving
\begin{equation}
    \hat{\bs{x}} = \argmin_{\bs{x}} \left\{ \frac{1}{2} \|\bs{r} - \bs{A}\bs{x} \|_2^2 + \tau \|\bs{x}\|_1 \right\},
\end{equation}
where $\bs{A}=[\bs{a}(\bs{\theta}_0),\ldots,\bs{a}(\bs{\theta}_{N-1})]\in\mathbb{C}^{M\times N}$ and $\tau>0$ controls sparsity. We adopt ADMM to decouple data fidelity and sparsity by introducing $\bs{z}\in\mathbb{C}^{N}$ and enforcing $\bs{x}=\bs{z}$:
\begin{equation}
    \min_{\bs{x}, \bs{z}} \quad \frac{1}{2} \|\bs{r} - \bs{A}\bs{x} \|_2^2 + \tau \|\bs{z}\|_1 \quad \text{s.t.} \quad \bs{x} = \bs{z}.
\end{equation}
The ADMM updates read
\begin{equation}
    \begin{aligned}
        \bs{x}^{(t+1)} &= (\bs{A}^\ctrans \bs{A} + \rho \bs{I} )^{-1} ( \bs{A}^\ctrans \bs{r} + \rho (\bs{z}^{(t)} - \bs{v}^{(t)}) ), \\
        \bs{z}^{(t+1)} &= \mathsf{R}_{\tau/\rho} ( \bs{x}^{(t+1)} + \bs{v}^{(t)} ), \\
        \bs{v}^{(t+1)} &= \bs{v}^{(t)} + \bs{x}^{(t+1)} - \bs{z}^{(t+1)},
    \end{aligned}
    \label{eqn:fixed_update_rules}
\end{equation}
with element-wise soft threshold $\mathsf{R}(0)=0$ and
\begin{equation}
    \mathsf{R}_{\tau/\rho}(\bs{u}) = \frac{\bs{u}}{|\bs{u}|} \max(0, |\bs{u}| - \frac{\tau}{\rho}).
    \label{eqn:fixed_threshold}
\end{equation}
In practice, fixed update rules \eqn \eqref{eqn:fixed_update_rules} and a fixed threshold \eqn \eqref{eqn:fixed_threshold} may over-shrink true components and slow convergence in coherent-source regimes, and the matrix inverse can be costly without structure \cite{o2016statistical}.

\noindent \textbf{Deep Unrolling}.
To overcome these issues while iteratively optimizing the spatial spectrum with ADMM, as shown in \fig \ref{fig:lego_beamformer}, we preserve the steering model and the classical input and output interface, and unroll the solver into a trainable network that learns only coefficients typically fixed in classical methods. Inspired by the iterative gated recurrent unit (GRU) \cite{chung2014empirical} architecture, we add a lightweight gate that connects $\bs{z}^{(t)}$ and $\bs{z}^{(t+1)}$, so each update blends the fresh candidate with the previous iterate. This acts as learned relaxation or momentum and is known to improve stability and speed in unrolled sparse solvers \cite{wu2020sparse}. We also replace the fixed soft-threshold with a learnable shrinkage level and parameterize step sizes to be positive via a softplus mapping, following evidence that learned thresholds and steps accelerate convergence and enhance robustness \cite{gregor2010learning,ito2019trainable}. 
Finally, a learnable diagonal preconditioner $\bs{W}^{(t)}$ keeps the linear solver inexpensive and provides data-adaptive weighting without resorting to subspace factorization \cite{boyd2011distributed}. 
The resulting solver removes reliance on brittle eigen-decompositions and, by learning an adaptive update strategy, achieves robust angular resolution even under challenging conditions.
At each iteration, the \sysname Beamformer updates read
\begin{equation}
    \label{eqn:rflego_bf}
    \begin{aligned}
        \bs{x}^{(t+1)} &= (\bs{W}^{(t)}+\bs{\eta}^{(t)}\bs{I})^{-1}
        (\bs{A}^{\ctrans}\bs{r}+\bs{\eta}^{(t)}(\bs{z}^{(t)}-\bs{v}^{(t)})),\\
        \bs{g}^{(t)} &= \sigma(\bs{W}_g\bs{z}^{(t)}+\bs{U}_g\bs{v}^{(t)}),\\
        \bs{z}^{(t+1)} &= \bs{g}^{(t)}\odot(\bs{x}^{(t+1)}+\bs{v}^{(t)})+(1-\bs{g}^{(t)})\odot\bs{z}^{(t)},\\
        \bs{v}^{(t+1)} &= \bs{v}^{(t)}+\bs{x}^{(t+1)}-\bs{z}^{(t+1)},
    \end{aligned}
\end{equation}
where $\bs{W}^{(t)}=\operatorname{diag}(\bs{w}^{(t)})$ is a learnable diagonal preconditioner, $\bs{\eta}^{(t)}=\operatorname{softplus}(\tilde{\bs{\eta}}^{(t)})>0$ enforces stable steps, and $\bs{g}^{(t)}\in[0,1]^N$ blends new and historical estimates.

\subsection{\sysname Detector}
\label{sec:sysname_dectector}

\begin{figure}[t]
    \centering
    \includegraphics[width=\linewidth]{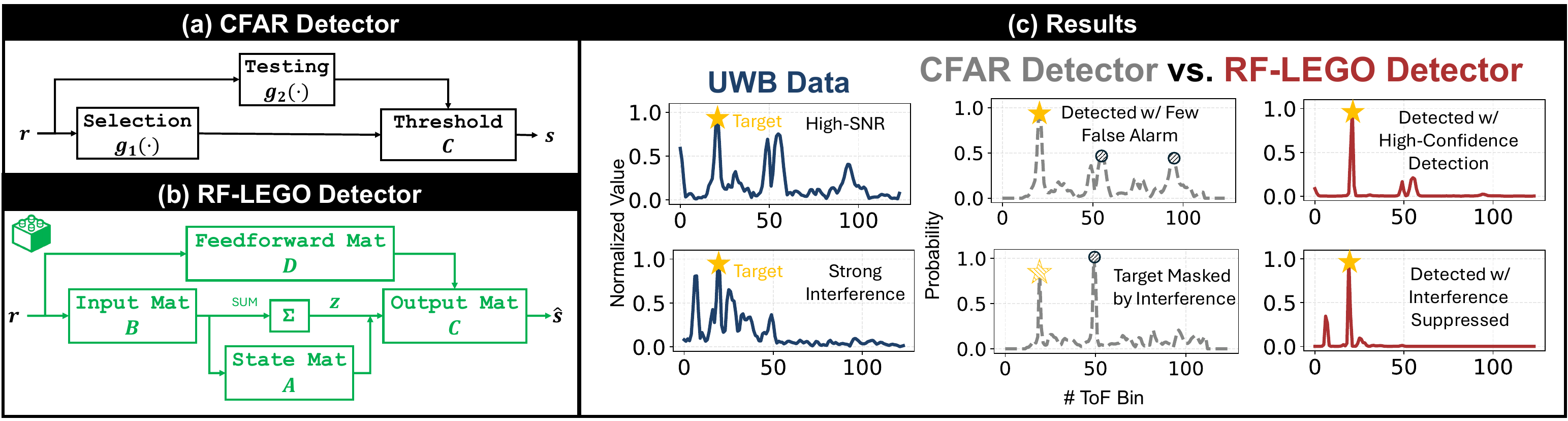}
    \caption{\sysname Detector. \rm{(a) Classical CFAR uses a fixed sliding window. (b) \sysname unrolls the logic into a state space model, where $\bs{r}$ is the input signal vector, $\bs{z}$ represents the learned state vector, and $\bs{\hat{s}}$ is the detection vector derived from the state.}}
    \label{fig:lego_detector}
\end{figure}

Detection underpins localization, tracking, and activity recognition. Classical matched detectors and Bayesian detectors \cite{van2004detection} require prior signal knowledge and degrade in dynamic, non-cooperative settings. 
Non-coherent energy and cyclostationary detectors (\eg, autocorrelation) offer alternatives but adapt poorly \cite{wright2022high}. 
CFAR and its variants, \ie, Cell Averaging (CA-CFAR) \cite{weiss2007analysis}, Order Statistic (OS-CFAR) \cite{rohling2007radar}, (Greatest Of) GO-CFAR \cite{Hansen1973}, and (Smallest Of) SO-CFAR \cite{truck1977range}, adjust thresholds using surrounding samples, yet still demand expert tuning and remain sensitive to heterogeneous clutter, limiting reliability in complex scenes. 
Recently, neural CFAR-like detectors report high accuracy with little manual tuning \cite{lin2019dl,tan2024dnn}, but they are often black boxes with limited physical relevance for RF sensing \cite{wright2022high} and are typically task-specific. 
There are already unrolled approaches for CFAR detection, such as CNN-based CFAR \cite{lin2019dl}, CNN-LSTM for maritime radar, OTFS-CFAR with neural components \cite{tan2024dnn}, CFARNet \cite{diskin2024cfarnet}, and VAMP-CFAR \cite{zhang2025parameter}. 
However, these prior works often face a critical dilemma: they either over-simplify the noise estimation model to ensure differentiability, which limits robustness in complex clutter, or they insert \emph{black-box} neural components that undermine the very interpretability that unrolling aims to preserve.

\noindent \textbf{Signal Model}.
The CFAR-based detection algorithms operate by systematically applying selection and hypothesis testing across a sliding window over the input signal $ \bs{r} \in \mathbb{R}^N$. At each step, a specific sample, denoted $ \overline{\bs{r}_n} $, is designated as the cell under test, while its surrounding samples, excluding a guard region, form a local neighborhood $\bs{r}_{guard}$. The detection output $ \bs{s} \in \mathbb{R}^N $ can then be expressed as:
\begin{equation}
    \small
    \bs{s}_n = \text{CFAR}(\bs{r}_n) = \bs{C} \bs{g}_1(\bs{r}_n) + \bs{g}_2(\bs{r}_n),
\end{equation}
where $ \bs{g}_1(\cdot) $ represents a selection operator that estimates the local noise level. The selection operator can be linear (\eg, CA-CFAR) or non-linear (\eg, OS-CFAR). 
$ \bs{g}_2(\cdot) $ serves as a testing operator that compares this estimate to the signal in the test cell. Typically, the testing operator takes a linear form, \ie, $\bs{g}_2(\bs{r}_n) = B\bs{r}_n$, where $\bs{B}$ is a linear transformation matrix and often simplifies to $\bs{B} \bs{r}_n = -\overline{\bs{r}_n}$. This formulation captures the essence of CFAR: a dynamically adapted thresholding process anchored in local statistics. Again, its performance is sensitive to the choice of parameters such as the threshold multiplier $\bs{C}$ and the configuration of the guard and training cells. These choices require manual tuning and are often non-trivial, particularly in complex environments where the improper configuration can significantly impair detection reliability.

\noindent \textbf{Deep Unrolling}. 
Our key insight to unroll CFAR is that it performs adaptive thresholding driven by temporally estimated noise, which aligns naturally with a state space model (SSM) architecture \cite{hamilton1994state, gu2021combining}. 
Moreover, an SSM offers minimal structured memory to track clutter or drift over time without altering the decision diagram of CFAR. We thus unroll CFAR as a discrete SSM, preserving the workflow and operating-point control while learning the noise/clutter dynamics.
We therefore unroll CFAR’s selection-testing routine into a discrete SSM, yielding a trainable detector with interpretability, which preserves the CFAR workflow.
Rather than applying the fixed operators $\bs{g}_1$ and $\bs{g}_2$ to the neighborhood samples $\bs{r}_{\text{guard}}$, we introduce a latent state $\bs{z}_n$ that evolves under learned dynamics. 
This hidden state will not affect the main information diagram but additionally aggregates higher-order statistics and other context and acts as an unobserved variable set, enriching the representation of the system’s internal behavior \cite{ling2010nonlinear,revach2022kalmannet}. 
\fig \ref{fig:lego_detector} shows the architecture: the CFAR workflow is preserved, while the matrices are trainable. 
The unrolled model is
\begin{equation}
    \bs{z}_n = \bs{A} \bs{z}_{n-1} + \bs{B} \bs{x}_n, \quad
    \bs{s}_n = \bs{C} \bs{z}_n + \bs{D} \bs{x}_n,
    \label{eqn:discrete-ssm}
\end{equation}
where $\bs{A}$ governs state evolution, $\bs{B}$ and $\bs{C}$ project inputs and states, and $\bs{D}$ is a feedforward term that plays the role of the CFAR testing operator $\bs{g}_2(\cdot)$. 
This formulation replaces the selection $\bs{g}_1(\cdot)$ with the learned state $\bs{z}_n$ and the fixed test with a linear readout over the state and current input. 
This discrete model can be viewed as the trapezoidal discretization of the continuous-time system:
\begin{equation}
    \frac{d \bs{z}(t)}{dt} = \bs{A} \bs{z}(t) + \bs{B} \bs{x}(t), \quad
    \bs{s}(t) = \bs{C} \bs{z}(t) + \bs{D} \bs{x}(t),
    \label{eqn:cont-ssm}
\end{equation}
which reduces to \eqn \eqref{eqn:discrete-ssm} by absorbing the step size $\Delta t$ into the learnable parameters. 
To numerically solve this ordinary differential equation, we adopt the Trapezoidal Rule to approximate the integral between steps $t_{n-1}$ and $t_n = t_{n-1} + \Delta t$:
\begin{equation}
    \begin{aligned}
        \bs{z}_n - \bs{z}_{n-1} &\approx \frac{\Delta t}{2} \left[
            \bs{A} \bs{z}_n + \bs{A} \bs{z}_{n-1} + \bs{B} \bs{x}_n + \bs{B} \bs{x}_{n-1}
        \right] \\
        \Rightarrow \bs{z}_n &= (\bs{I} - \tfrac{\Delta t}{2} \bs{A})^{-1} \left[(\bs{I} + \tfrac{\Delta t}{2} \bs{A}) \bs{z}_{n-1} + \Delta t\, \bs{B} \bs{x}_n \right],
        \label{eqn:trapezoid}
    \end{aligned}
\end{equation}
which simplifies to the form in \eqn \eqref{eqn:discrete-ssm} by absorbing the time-step $\Delta t$ into the learnable parameters. Therefore, our network preserves physical fidelity to the underlying continuous dynamics while allowing flexible, data-driven adaptation via trainable matrices $\bs{A}, \bs{B}, \bs{C}, \bs{D}$. 
As illustrated in \fig \ref{fig:lego_detector}, the design preserves the CFAR structure of selection and testing, adds memory to store latent information, and introduces trainable components. In practice, the detector adapts to environmental variation, generalizes better across unseen scenarios, and remains interpretable because each component maps to a clear physical role.

\subsection{Remarks}
\label{sec:remarks}
\sysname pioneers deep unrolled SP-DL co-design for RF sensing. 
We present three unrolled algorithms for the most fundamental processing in sensing applications, \ie, frequency transform, spatial spectrum estimation, and detection. 
The three proposed modules can be used separately or combined together, like LEGO blocks. For example, one can perform \sysname FT and then \sysname Detector for breathing rate estimation. 
They can also be flexibly reused and integrated into prior and emerging sensing solutions, since the unrolled modules preserve a similar interface, just like the classical signal processing techniques. 
We leave it as future work to unroll the above three techniques in different ways and unroll other important techniques in RF sensing. 

\begin{figure*}[t]
    \centering
    \includegraphics[width=\linewidth]{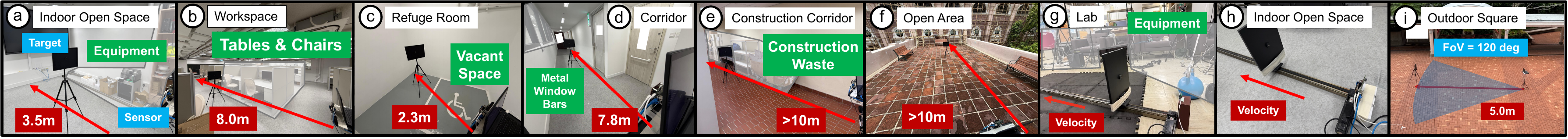}
    \caption{Experimental scenarios. \rm{(a-f) Range experiments; (g-h) Doppler experiments; (g,i) Angle experiments.}}
    \label{fig:rflego_scenario}
\end{figure*}

\section{Implementation}
\label{sec:implementation}
A key advantage is that each \sysname module is trained on synthetic data, yet, as our experiments show, it generalizes robustly to diverse real-world scenarios. We synthesize 30k frames for each module that mimic real measurements while exposing controlled variability.

\noindent \textbf{\sysname FT}.
We first construct clean frequency-domain spectra, then inject typical artifacts (\eg, spectral leakage) and additive white Gaussian noise across 5-40 dB SNR. An inverse Fourier Transformation converts these spectra to noisy time-domain signals for training. The default frequency transform points are 256.

\noindent \textbf{\sysname Beamformer}.
We simulate a uniform linear array with a steering dictionary. We draw a random set of sources with random directions and complex amplitudes to form a clean array snapshot, then add white Gaussian noise. The ground truth is a sparse vector on a discrete angle grid whose nonzero entries correspond to the dictionary indices nearest the true directions. By default, the array has 8 antennas, 10 ADMM layers, and the steering grid spans from -60 to 60 degrees at 1-degree resolution.

\noindent \textbf{\sysname Detector}.
We synthesize one-dimensional signals by superimposing 1-5 target peaks on unit-variance white Gaussian noise. Peaks use Hann or Hamming shapes with varying widths and are scaled to achieve 5-40 dB SNR. Labels are binary masks marking the peak locations. Each sample has a length of 128.

All modules are implemented in PyTorch and trained on a single NVIDIA GeForce RTX 4090. We use AdamW with a learning rate $1\!\times\!10^{-3}$ and a weight decay of 0.01, batch size 512, and an 80/20 train-validation split; dropout of 0.2 is applied within modules. Losses are task-aligned: cosine similarity for \sysname FT and \sysname Beamformer, and binary cross-entropy for \sysname Detector. Computation/size costs are: \sysname FT (0.02 GFLOPs, 0.8M), Beamformer (0.83 GFLOPs, 0.6M), and Detector (0.05 GFLOPs, 0.4M).

\section{Evaluation}
\label{sec:evaluation}
Our evaluation benchmarks individual modules against signal processing, deep learning, and loose coupling baselines, validates plug-and-play cascadability in existing pipelines, and analyzes structure-aligned interpretability.
\subsection{Experiment Design}
\subsubsection{Experiment Setup}
To evaluate the modularity of \sysname across diverse RF-based platforms, we utilize a range of IoT devices, including mmWave radar, UWB radar, and Wi-Fi. Evaluation experiments are tailored to the distinct features of each sensor's electromagnetic properties and modulation techniques.

\begin{itemize}[noitemsep, topsep=0pt, leftmargin=*]
    \item \textbf{mmWave}. IQ-modulated FMCW raw signals are captured from the TI IWR1843 \cite{TI_IWR1843BOOST} using the DCA1000EVM evaluation board \cite{TI_DCA1000EVM}, operating within the 77-81 GHz band.
    
    \item \textbf{UWB}. Impulse-based time-of-flight (ToF) signals are collected using the Novelda XeThru X4A02 ultra-wideband radar \cite{novelda_x4_user_guide}, which operates at a center frequency of 7.29 GHz and a frame rate of 100 Hz.
    
    \item \textbf{Wi-Fi}. Wi-Fi signals are acquired using a commercial router and an Intel AX200 NIC. The transmitter antennas send packets to two receiver antennas, enabling the extraction of channel state information (CSI).
\end{itemize}
To accurately benchmark our algorithms, we selected a metal plate as an ideal point-like reflector. 
Its large radar cross-section not only ensures a strong and unambiguous echo for simplified ground truth acquisition but also minimizes confounding variables arising from complex target geometries. 
As \fig \ref{fig:rflego_scenario} shows, we design distinct, controlled experiments across nine scenarios for the three key sensing dimensions:
\head{Range} The target is positioned at multiple distances from the sensor to capture distance-dependent signal returns.
\head{Doppler} The plate is mounted on a programmable linear rail and driven at precisely controlled velocities, with the sensor fixed at one end of the rail to record the resulting frequency shifts.
\head{Angle} To evaluate performance at varying distances, the target is placed on circular tracks with 3 m and 5 m radii (corresponding to scenarios g and i, respectively) and rotated through a sweep of azimuth angles relative to the sensor.
For each modality and each scenario, we collect approximately 36k samples spanning range, Doppler, and angle, and every sequence is labeled with its exact measurement.

\subsubsection{Evaluation Metrics}
To quantify the performance of the three \sysname modules, we use a suite of metrics that assess their core capabilities in spectral analysis, parameter estimation, and target detection.

\begin{figure*}[t]
    \centering
    \includegraphics[width=\linewidth]{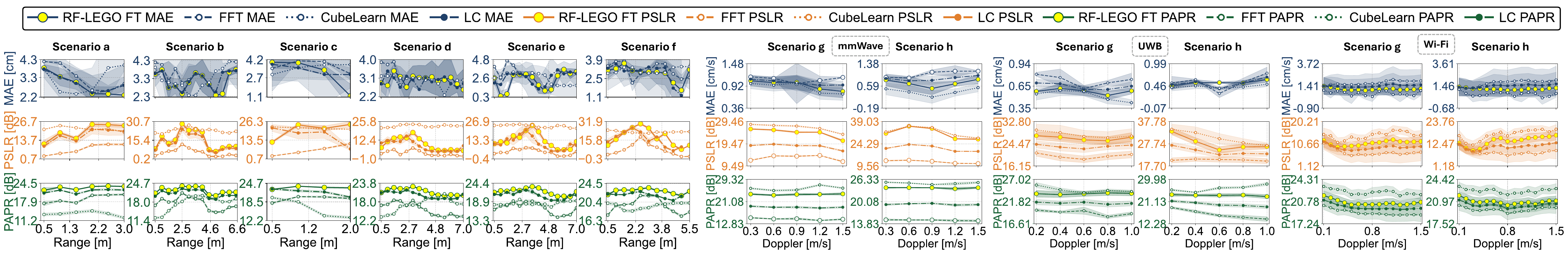}
    \caption{The results of  \sysname Range FT of mmWave and  \sysname Doppler FT of mmWave, UWB, and Wi-Fi.}
    \label{fig:rflego_ft_results}
\end{figure*}

\begin{figure*}[t]
  \begin{minipage}{0.31\textwidth}
    \centering
    \includegraphics[width=\linewidth]{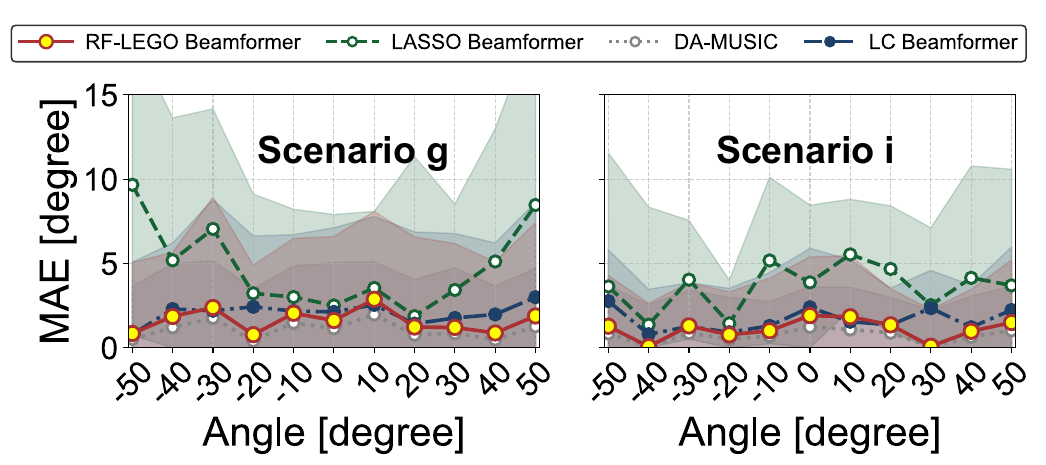}
    \caption{The results of \sysname Beamformer of mmWave.}
    \label{fig:rflego_beamformer_results}
  \end{minipage}
  \hfill
  \begin{minipage}{0.32\textwidth}
    \centering
    \includegraphics[width=\linewidth]{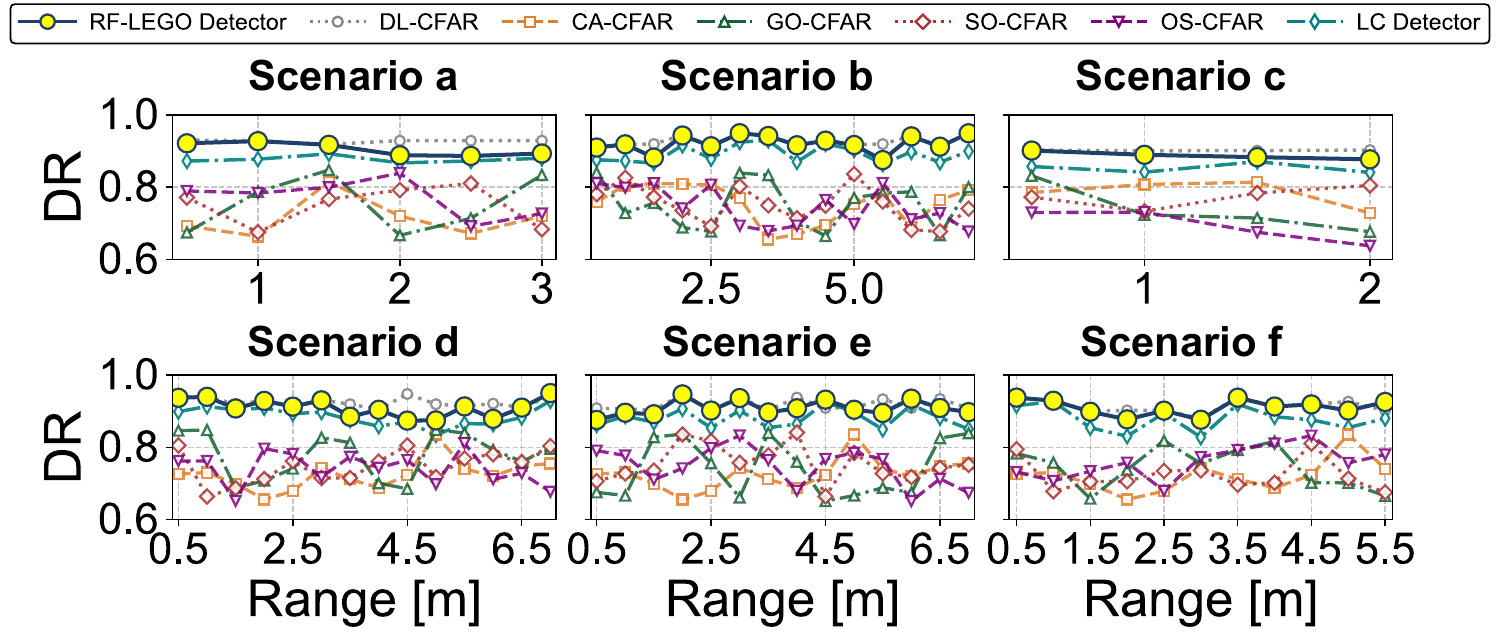}
    \caption{The results of  \sysname Detector of UWB ToF signals.}
    \label{fig:rflego_detector_results}
  \end{minipage}
  \hfill
  \begin{minipage}{0.36\textwidth}
    \centering
    \includegraphics[width=\linewidth]{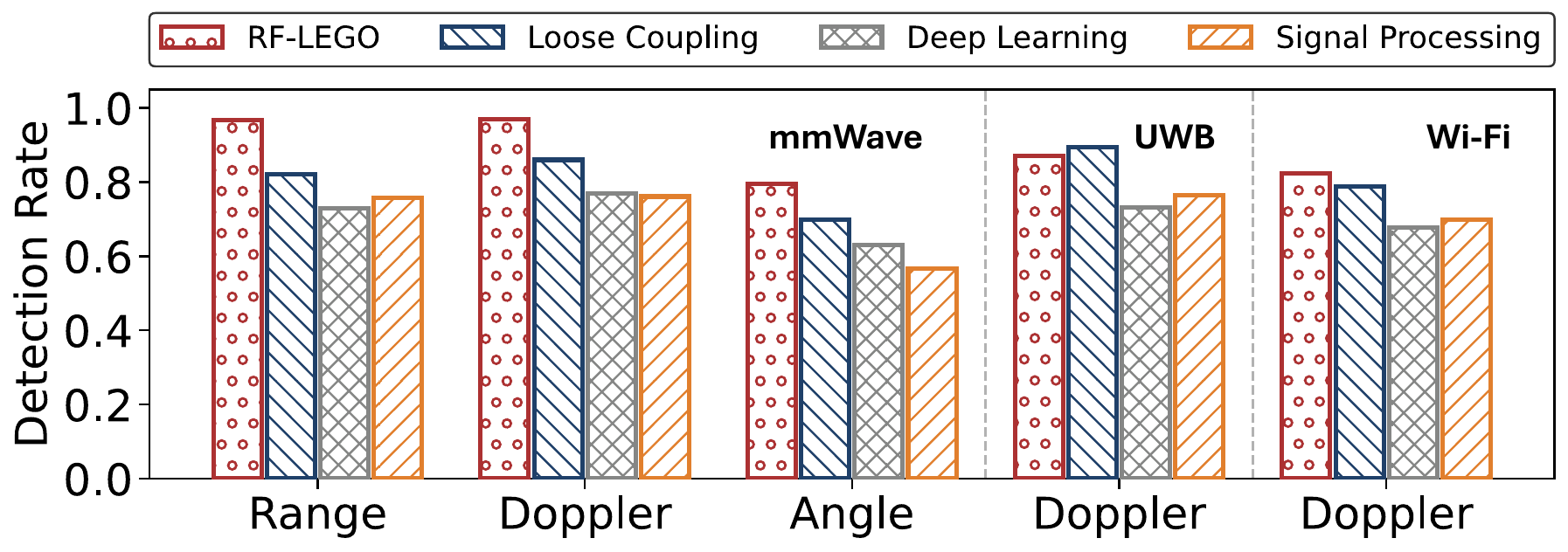}
    \caption{Cascadability evaluation of \sysname \vs  SP and LC baselines.}
    \label{fig:cascadability_evaluation_results}
  \end{minipage}
\end{figure*}

\head{Peak-to-Side Lobe Ratio (PSLR)} Measures the ratio between the main lobe's peak and the highest side lobe, quantifying spectral leakage suppression. A higher value indicates better isolation of strong signals from weaker adjacent.

\head{Peak-to-Average Power Ratio (PAPR)} Evaluates spectrum sparsity by comparing the peak power to the average power. A higher PAPR implies more concentrated signal energy and a cleaner, less cluttered spectrum.

\head{Mean Absolute Error (MAE)} Computes the absolute error between the estimated target and its ground truth value. A lower MAE indicates higher estimation accuracy for target parameters such as range, Doppler, or angle.

\head{Detection Rate (DR)} Represents the percentage of correctly identified targets, reflecting the model's ability to resolve distinct objects.

\subsubsection{Baselines}
We compare \sysname against various signal processing, deep learning and loose coupling baselines.

\begin{itemize}[noitemsep, topsep=0pt, leftmargin=*]
    \item \textbf{Signal Processing Baselines}: We compare our three modules against their most fundamental and direct signal processing counterparts: the conventional FFT, the iterative LASSO beamformer, and the CFAR detector, respectively.

    \item \textbf{Deep Learning Baselines}: 

        \head{(i) CubeLearn} \cite{zhao2023cubelearn}
        Represents the purely data-driven approach, which learns features directly from raw signals. For a fair comparison, we adapt it as a learned frequency transform front-end by removing its final classifier.
        
        \head{(ii) DA-MUSIC} \cite{merkofer2023music}
        Represents a hybrid approach that integrates a deep neural network into the classical MUSIC pipeline to improve beamforming, serving as a key baseline for data-driven angle estimation.
    
        \head{(iii) DL-CFAR} \cite{lin2019dl}
        Represents the component-replacement approach, where the noise estimation stage of classical CFAR is replaced with a ResNet-based module, enhancing a specific component in a purely neural network manner.

    \item \textbf{Loose Coupling Baselines (LC)}: We cascade a fixed SP front-end with a deep learning back-end, where the neural layers are configured to match the parameter count of the corresponding \sysname modules for a fair comparison.

\end{itemize}

\subsection{Experimental Evaluation}
\label{sec:experimental_evaluation}

\subsubsection{Modularity Evaluation}
To validate the modularity of \sysname, we conduct a series of rigorous module-swap control experiments. In this paradigm, we keep the classical signal processing pipeline intact, replacing only a single block with its \sysname counterpart to cleanly isolate and quantify the contribution of each component. 

First, as \fig \ref{fig:rflego_ft_results} illustrates, the \sysname FT module demonstrates a marked performance improvement over the conventional FT in frequency transformation tasks across all tested modalities. By suppressing spectral leakage through its learned, data-driven filter, the module achieves an average PSLR of around 24 dB and 27 dB and PAPR of approximately 16 dB and 23 dB for range and Doppler spectra, respectively. 
This superior spectral quality directly translates into more robust peak detection, significantly reducing range and velocity estimation errors (MAE). As shown in \fig \ref{fig:rflego_ft_results}, this corresponds to an average MAE reduction of approximately 10\% for range and 20\% for Doppler compared to the signal processing baseline.
Compared with the deep learning baseline, CubeLearn \cite{zhao2023cubelearn}, \sysname FT achieves a comparable accuracy, while offering the distinctive interpretability.
And it consistently outperforms the loose coupling baseline, due to the structured SP-DL co-design.
Subsequently, the \sysname Beamformer demonstrates a significant leap in spatial angle estimation over the classical LASSO baseline, as in \fig \ref{fig:rflego_beamformer_results}. On mmWave signals across scenarios g and i, it reduces MAE from 4.23 to 1.35 degrees by 68\%. 
Although DA-MUSIC \cite{merkofer2023music} also offers similar accuracy, \sysname achieves this competitive performance with two key advantages. First, it retains pipeline interpretability by outputting a physically meaningful angular spectrum. Second, its ADMM-based structure avoids the reliance on eigenvalue decomposition, an operation central to the MUSIC algorithm. This is a critical design choice, as eigenvalue decomposition is known to have unstable gradients within deep learning frameworks, often complicating or destabilizing the training process for models like DA-MUSIC \cite{pytorch2024eig}. 
Despite the strong performance of the LC baseline, \sysname Beamformer still achieves a lower mean angular error.
Finally, \fig \ref{fig:rflego_detector_results} shows the performance of \sysname Detector for target detection using UWB ToF signals. 
We compare \sysname Detector against both the traditional CFAR family and the end-to-end DL-CFAR model \cite{lin2019dl} on the UWB dataset. At a matched false alarm rate of $10^{-3}$, our \sysname Detector achieves a considerably higher DR than classical CFAR and a competitive performance compared to the DL-CFAR baseline. 
Crucially, \sysname Detector allows for explicit control over the detector's operating points, \eg, the trade-off between detection and false alarm rates, an important feature for building practical systems in real-world RF sensing. 
Moreover, \sysname Detector performs on par with the loose coupling baseline even under challenging conditions, while maintaining the co-designed SP structure.
In summary, these modularity evaluations prove \sysname as drop-in modules that consistently surpass its classical SP counterparts, the loose coupling baseline, and prior DL methods across different metrics.

\subsubsection{Cascadability Evaluation}
We also demonstrate that \sysname components are cascadable, composing seamlessly into high-performance pipelines with compounding benefits.
We test this by constructing two hybrid pipelines: (i) \sysname FT plus \sysname Detector for range/velocity estimation, and (ii) \sysname Beamformer plus \sysname Detector for angle estimation.
We benchmark these against classical pipelines, cascaded loose-coupling pipelines, and cascaded end-to-end baselines.
We use the Detection Rate (DR) of the ground truth target bin at a fixed false alarm rate of $10^{-3}$ as the primary metric.
As shown in \fig\ref{fig:cascadability_evaluation_results}, the cascaded \sysname pipelines achieve higher DR than classical pipelines, with improvements of 27.6\% (range), 27.3\% (Doppler), and 40.2\% (angle) on mmWave data, and further gains of an average of 28.2\% and 14.7\% over the cascaded deep learning and loose coupling baselines.
Similar advantages are observed on UWB and Wi-Fi data. 
Results confirm \sysname outperforms standalone methods while enabling reusable, cascadable complex sensing.

\subsubsection{Interpretability Analysis}
\label{sec:interpretability_analysis}
\sysname ensures structure-aligned interpretability through the SP-structured module design. 
This structure interpretability allows each trainable module to be directly inspected and reasoned about using principled, SP-style analysis of intermediate results with classical criteria and metrics. 
For instance, we can calculate PSLR/PAPR for \sysname FT outputs in \fig \ref{fig:rflego_ft_results}, quantifying spectral quality (\eg, sidelobes) in the same way commonly used in conventional spectral analysis via signal processing. 
Moreover, the structure-aligned interpretability supports flexible and robust cascading when composing multi-module pipelines. Thanks to the interpretable structures with trainable modules, \sysname-enhanced pipelines are more accurate and more resilient to inter-module error propagation, as demonstrated in \fig \ref{fig:cascadability_evaluation_results} and \fig\ref{fig:interface_sensitivity_under_cascading}.

\subsection{Microbenchmarks}
\begin{figure*}[t]
    \centering
    \includegraphics[width=\linewidth]{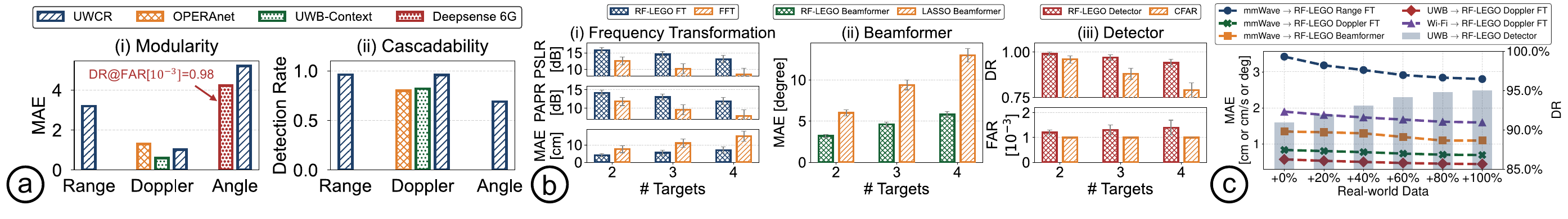}
    \caption{Microbenchmarks. \rm{(a) Performance on public datasets. (b) Impact of multiple targets. (c) Impact of data fine-tuning.}}
    \label{fig:micro_benchmarks_results}
\end{figure*}

\subsubsection{Performance on Public Datasets}
We evaluate \sysname across multiple online datasets:

\noindent \rev{\textbf{UWCR}} \cite{9249018,xm40-jx59-22}: The dataset contains real urban and parking-lot driving scenes with synchronized radar-camera data. We use it to assess \sysname Range FT, Doppler FT, and Beamformer. The ground truth is provided by radar-camera fusion annotations derived from vision labels aligned to radar frames with manual verification.

\noindent \textbf{OPERAnet} \cite{bocus2022operanet}: The dataset offers indoor Wi-Fi CSI with synchronized Kinect and camera data. We align skeleton trajectories to the sensor frame to obtain radial-velocity ground truth and evaluate \sysname Doppler FT. 
    
\noindent \textbf{UWB-Context} \cite{bocus2022comprehensive}: The dataset contains residential multi-static UWB CIRs. We convert CIR to ToF to evaluate \sysname Doppler FT and Detector. Ground truth 2D positions are provided by a synchronized, independent active UWB.

\rev{\noindent \textbf{DeepSense 6G} \cite{alkhateeb2023deepsense}: The dataset features large-scale real-world 60 GHz millimeter-wave communication measurements, \ie, 6G, collected via a phased array by beam sweeping. We utilize the received power vectors obtained during the scan to assess \sysname Detector for robust beam selection tasks. The ground truth is determined by mmWave radar.}

\begin{table}[t]
    \centering
    \caption{Ablation study.}
    \includegraphics[width=\linewidth]{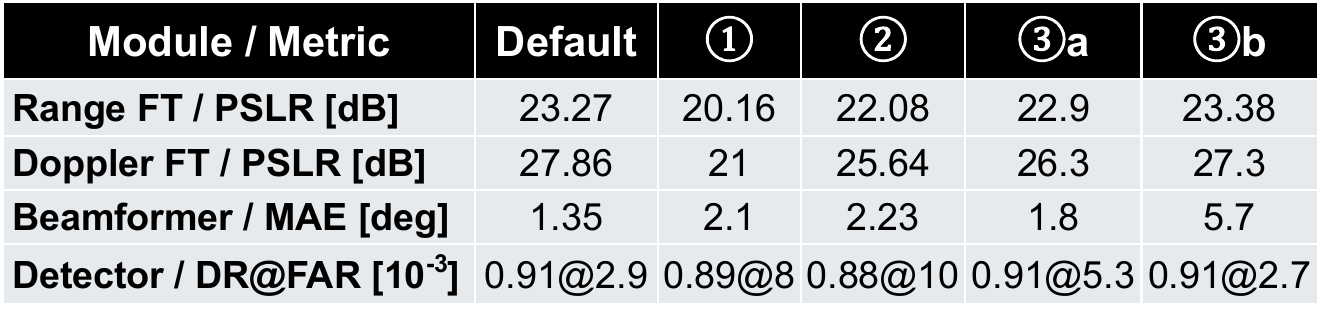}
    \label{tab:ablation_study}
\end{table}

\subsubsection{Impact of Data Fine-tuning}
By default, \sysname modules are trained exclusively on synthetic data. For our main experiments, these pre-trained models are then directly evaluated on real-world data without any fine-tuning.
In this experiment, we fine-tune them with a varying fraction of real data per modality from 0-100\%. As shown in \fig\ref{fig:micro_benchmarks_results}(c), fine-tuning with real-data further enhances \sysname: the MAE decreases for Range FT (about 18\% from 0\% to 100\%), Doppler FT on mmWave, UWB, and Wi-Fi (about 15-25\%), and the Beamformer (about 15\%), while the DR of Detector rises by about 5-6\% at a fixed order of magnitude of FAR. The gains are monotonic and begin to plateau after roughly 60-80\%, indicating effective low-shot adaptation with diminishing returns at larger fractions. This suggests a practical deployment workflow where a small amount of in-domain data can be used for finetuning to recover most gains.

\subsubsection{Impact of Multiple Targets}
To assess the effect of multiple targets, we conduct mmWave experiments with 2-4 targets placed at random ranges of 1-4 m and random azimuths of -60 to 60 degrees. We collect approximately 8k samples across 27 configurations. As \fig\ref{fig:micro_benchmarks_results}(b) shows, \sysname consistently outperforms classical signal processing baselines across all target counts and metrics, indicating robustness to mutual interference and clutter without fine-tuning.

\subsubsection{Interface Sensitivity under Cascading}
\label{sec:interface_sensitivity_under_cascading}
\begin{figure}[t]
    \centering
    \includegraphics[width=\linewidth]{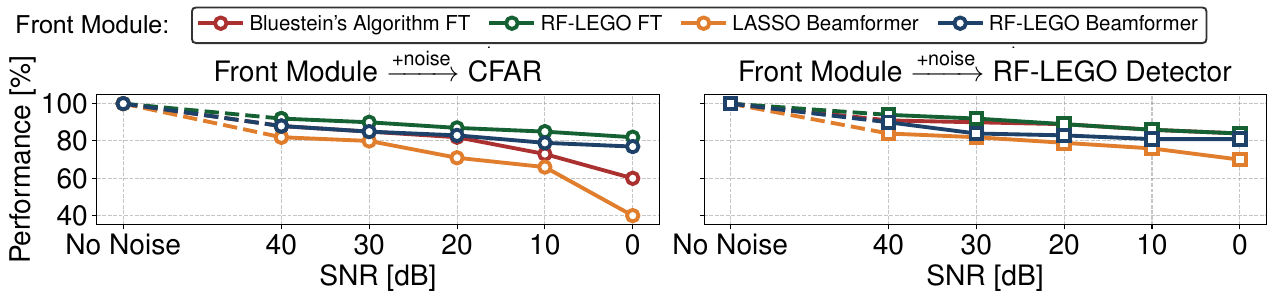}
    \caption{Effect of interface sensitivity under cascading.}
    \label{fig:interface_sensitivity_under_cascading}
\end{figure}

\fig \ref{fig:interface_sensitivity_under_cascading} evaluates error propagation by injecting noise of different SNR levels between cascaded mmWave sensing modules and measuring the normalized detection rate. As expected, noise degrades performance for all pipelines; Nevertheless, \sysname consistently exhibits slower performance decay than pure signal processing pipelines, indicating stronger resilience to upstream errors. Moreover, robustness improves as more \sysname modules are integrated, suggesting that the unrolled, physics-constrained operators remain compatible under cascading and mitigate cumulative error amplification. Overall, these results confirm that \sysname’s cascadability is not merely due to downstream tolerance, but stems from enhanced error resilience at each unrolled module.

\subsubsection{Inference Latency on Edge Devices}

We also report the inference latency breakdown of each \sysname module on representative edge platforms under default settings. As detailed in \tab \ref{tab:edge_devices}, \sysname achieves fast performance on the Jetson Orin Nano by incurring additional computation while improving task performance. While it remains practical on the Raspberry Pi 4, the microcontroller-class ESP32-P4 presents the most significant constraints. For practical deployment, effective model compression can be further performed for extremely resource-constrained hardware.

\begin{table}[t]
    \centering
    \caption{Inference latency on different edge platforms. \rm{The results are formatted as \sysname / SP [ms].}}
    \includegraphics[width=\linewidth]{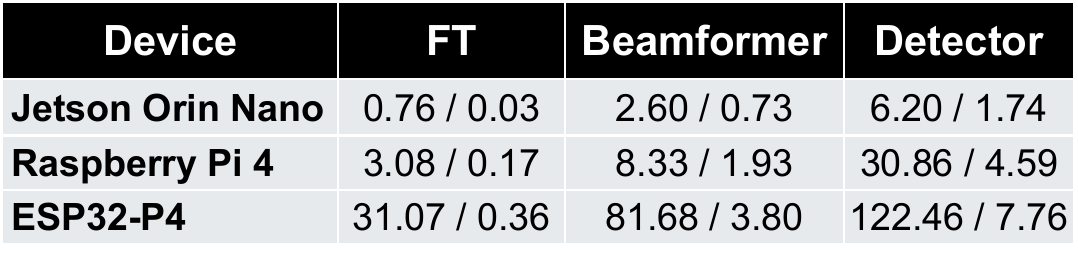}
    \label{tab:edge_devices}
\end{table}

\subsection{Ablation Study}
\label{sec:ablation_study}

We evaluate the unrolled design of each module, and the results are depicted in \tab \ref{tab:ablation_study}.

\noindent\textbf{\sysname FT}\;
\ding{172} \emph{w/o activation}: removing nonlinearities degrades PSLR by 3.1 dB and 6.9 dB for Range and Doppler FT, respectively, confirming that mild nonlinearity suppresses leakage beyond pure signal processing convolution. 
\ding{173} \emph{w/o nulling}: performance is comparable to the main model, indicating gains mainly come from the unrolled architecture rather than the nulling head.
\ding{174} \emph{\# points}: using 128 (a) or 512 (b) points is on par with the default 256, showing good generalizability.

\noindent\textbf{\sysname Beamformer}\;
\ding{172} \emph{w/o iteration connection}: removing the explicit link to the previous iterate increases MAE (from 1.35 to 2.10 degrees), highlighting the benefit of iterative memory under challenging conditions.
\ding{173} \emph{w/o gating}: replacing the learnable gate with a fixed soft-threshold further worsens MAE to 2.23 degrees, showing that the data-adaptive step is key to robustness.
\ding{174} \emph{\# antennas}: reducing the ULA from 8 to (a) 6 causes moderate degradation, while reducing to (b) 4 sharply degrades accuracy to 5.7 degrees, matching the expected resolution loss.

\noindent\textbf{\sysname Detector}\;
\ding{172} \emph{w/o activation}: detection rate drops slightly while false alarm rate increased by approximately 2.8$\times$, indicating that nonlinear capacity helps suppress false alarms in co-design. 
\ding{173} \emph{w/o trainable state}: setting the state matrix to a fixed identity matrix, \ie, $\bs{A}=\bs{I}$, reduces DR to 0.88 and raises FAR by an order of magnitude, showing the state stores useful latent information.
\ding{174} \emph{input length}: using (a) 64 or (b) 256 bins yields results close to default (128).

\subsection{Module Behavior Analysis}
\label{sec:module_behavior_anaylsis}

\begin{figure}[t]
    \centering
    \includegraphics[width=\linewidth]{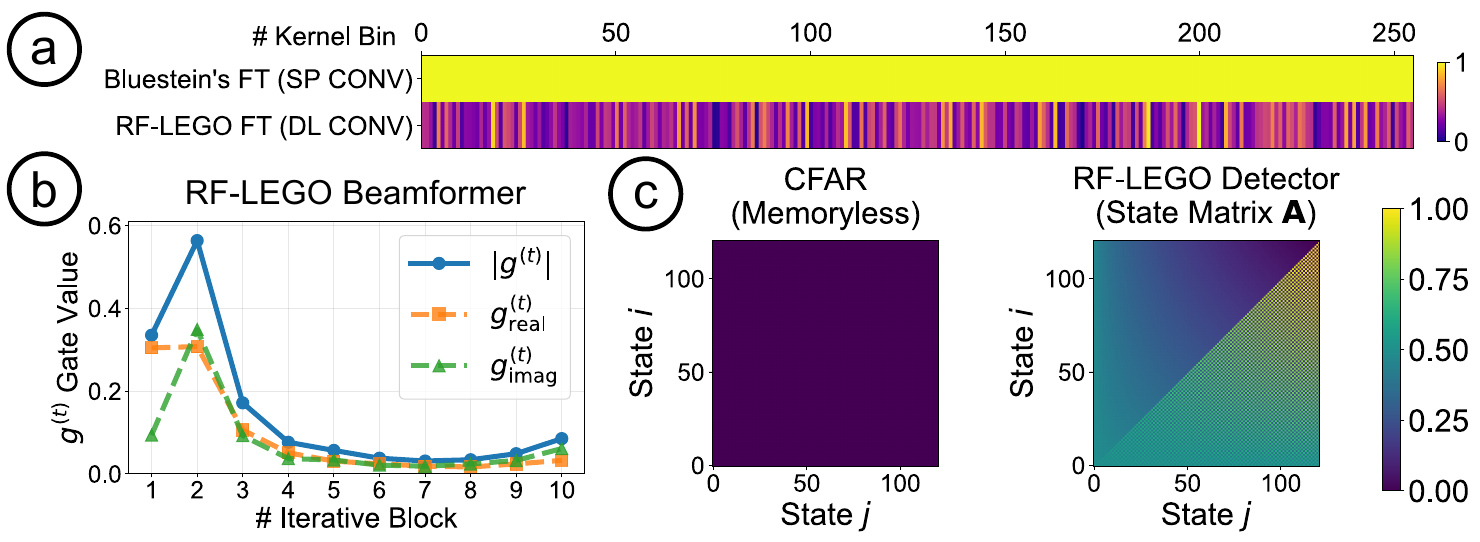}
    \caption{Module behavior analysis. \rm{(a) \sysname FT learns a non-uniform convolution kernel beyond Bluestein's Algorithm FT. (b) \sysname Beamformer learns adaptive gate schedules across unrolled iterations. (c) \sysname Detector learns a structured state matrix $\bs{A}$, unlike memoryless CFAR.}}
    \label{fig:module_analysis}
\end{figure}

\begin{figure*}[t]
  \begin{minipage}{0.49\textwidth}
    \centering
    \includegraphics[width=\linewidth]{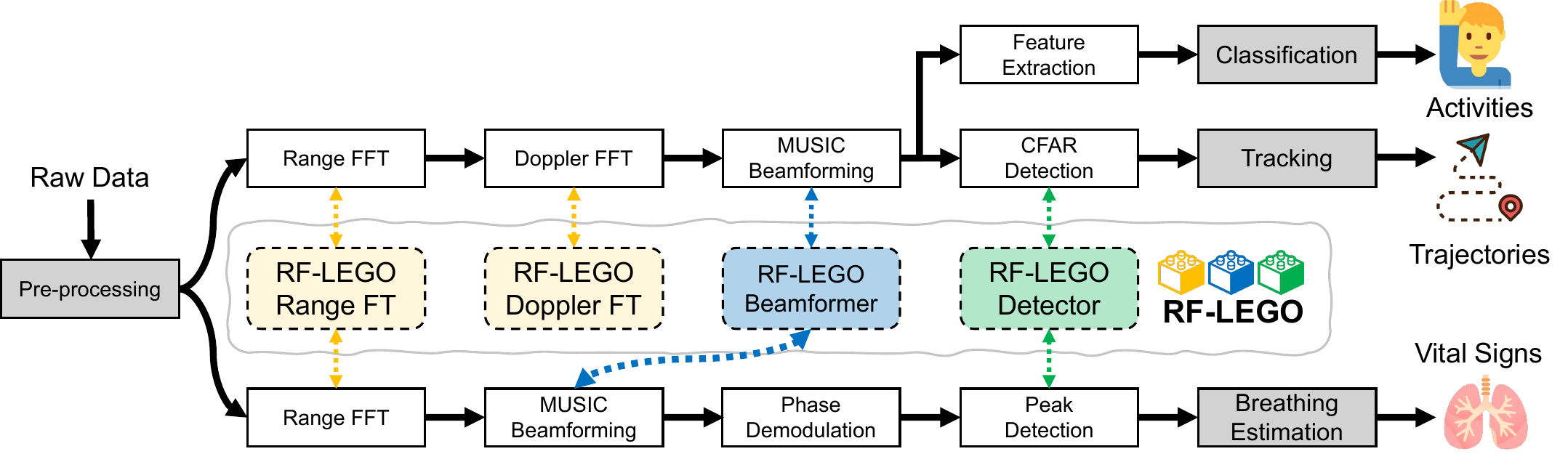}
    \caption{Pipelines for trajectory tracking, vital sign monitoring, and human activity recognition with \sysname modules \vs classical baselines.}
    \label{fig:case_study_pipeline}
  \end{minipage}
  \hfill
  \begin{minipage}{0.49\textwidth}
  \centering
    \includegraphics[width=\linewidth]{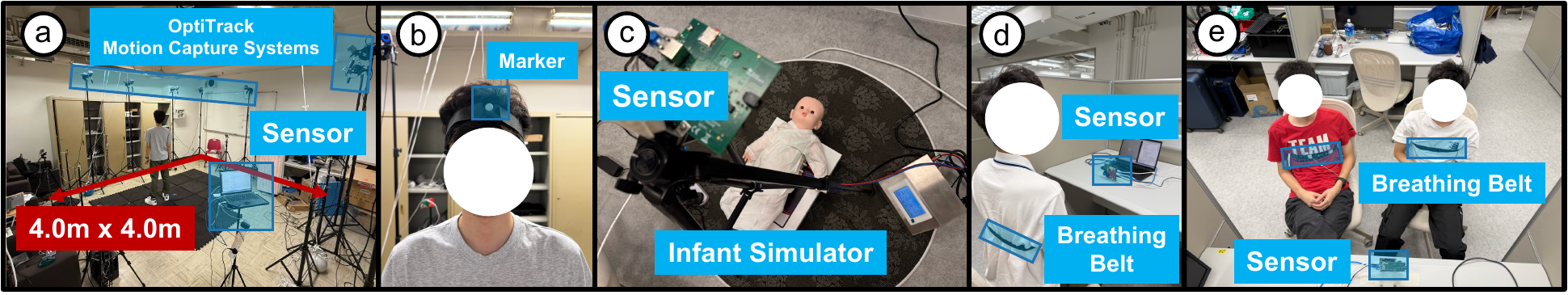}
    \caption{Case study scenarios. \rm{(a-b) Trajectory tracking; (c-e) Vital sign monitoring for infant simulator and human adults.}}
    \label{fig:case_study_scenario}
  \end{minipage}
\end{figure*}

Compared with Bluestein’s Algorithm FT, whose signal processing convolution kernel has a uniform response across kernel bins, the \sysname FT learns a data-driven convolution kernel. As shown in \fig \ref{fig:module_analysis}(a), this manifests as a clear reweighting pattern over kernel bins, enabling the FT module to adapt the intermediate convolution to non-ideal measurements and thus outperform the fixed-kernel baseline in downstream performance. \fig \ref{fig:module_analysis}(b) shows the evolution of the learned gate $g^{(t)}$ across unrolled iterative blocks. The gate values indicate an iteration-dependent focus between the previous iterate and the newly-updated estimate in the update candidate, \ie, $\bs{x}^{(t+1)}+\bs{v}^{(t)}$ in \eqn \eqref{eqn:rflego_bf}. Specifically, larger gates in early blocks encourage stronger incorporation of the update candidate, while smaller gates in later blocks emphasize stability, reflecting a learned coarse-to-fine scheduling that improves convergence behavior. \fig \ref{fig:module_analysis}(c) compares the learned state matrix $\bs{A}$ with the memoryless CFAR baseline. While CFAR corresponds to a null state transition, \sysname Detector learns a structured $\bs{A}$ and meanwhile maintains the stable structural pattern inherited from its initialization. This is desirable because it preserves a constrained state evolution while still allowing data-driven adaptation, yielding a detector with memory modeling beyond local statistics.

\section{Case Study}
\label{sec:case_study}
We evaluate \sysname's impact by replacing SP modules in three pipelines, as shown in \fig \ref{fig:case_study_pipeline}, while keeping downstream components fixed. For tracking, \sysname replaces the baseline's \cite{gupta2019openradar} FFT, MUSIC, and CFAR feeding an extended Kalman filter. For vital signs, we compare breathing rates against \cite{ti2025vitalsigns} using identical phase analysis. For activity recognition \cite{9894724}, \sysname replaces the SP front-end of the baseline \cite{10001175} classifier.

\begin{figure*}[t]
  \begin{minipage}{0.32\textwidth}
    \centering
    \includegraphics[width=\linewidth]{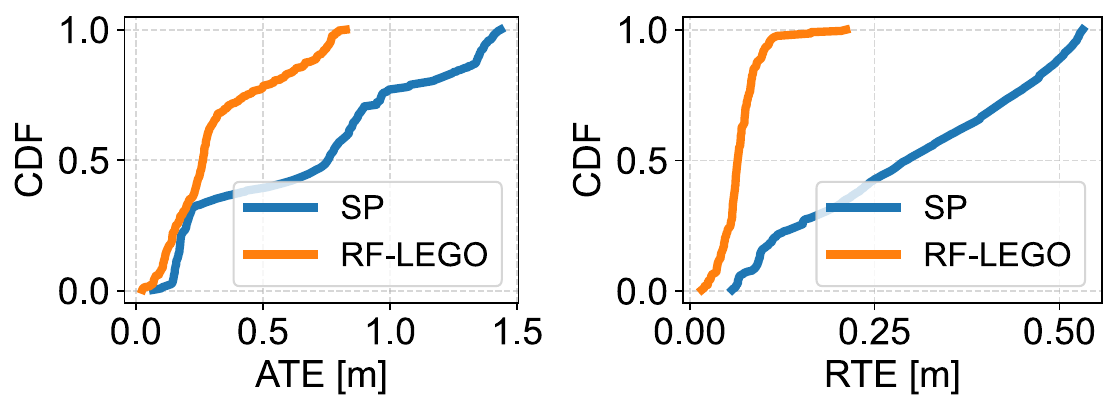}
    \caption{CDF of ATE and RTE.}
    \label{fig:cdf_of_ate_rte}
  \end{minipage}
  \begin{minipage}{0.5\textwidth}
  \centering
    \includegraphics[width=\linewidth]{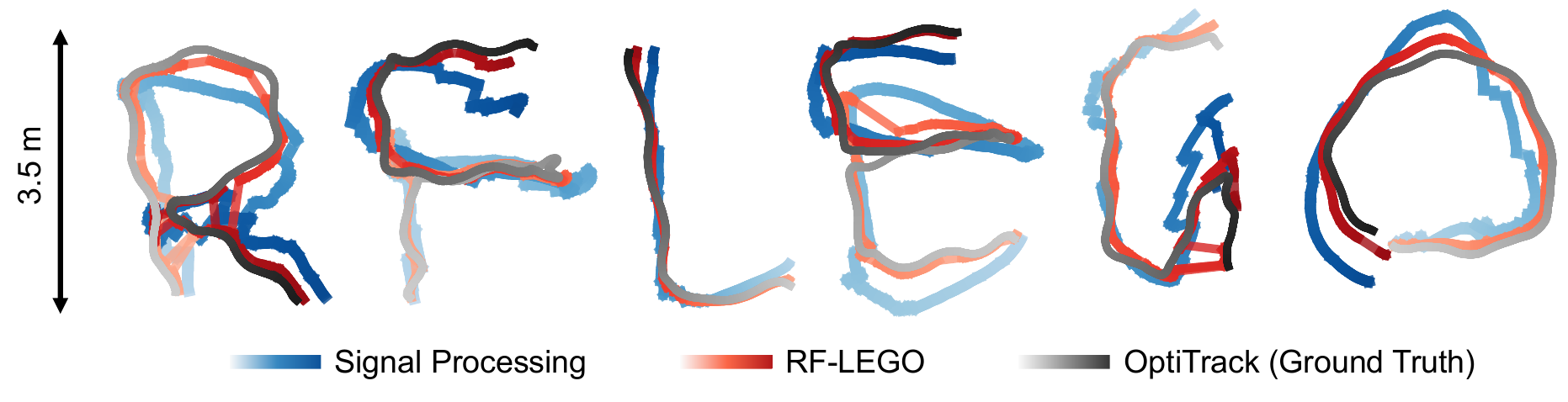}
    \caption{Human trajectory tracking for letters \sysname. \\ \rm{Colors fade from light (start) to dark (end).}}
    \label{fig:case_study_trajectory_sample}
  \end{minipage}
  \begin{minipage}{0.16\textwidth}
    \centering
    \includegraphics[width=\linewidth]{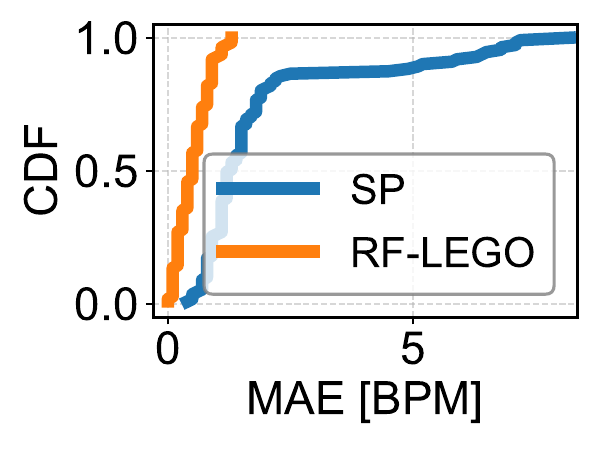}
    \caption{CDF of breathing MAE.}
    \label{fig:case_study_breathing_cdf}
  \end{minipage}
\end{figure*}

\begin{figure*}[t]
  \begin{minipage}{0.33\textwidth}
    \centering
    \includegraphics[width=\linewidth]{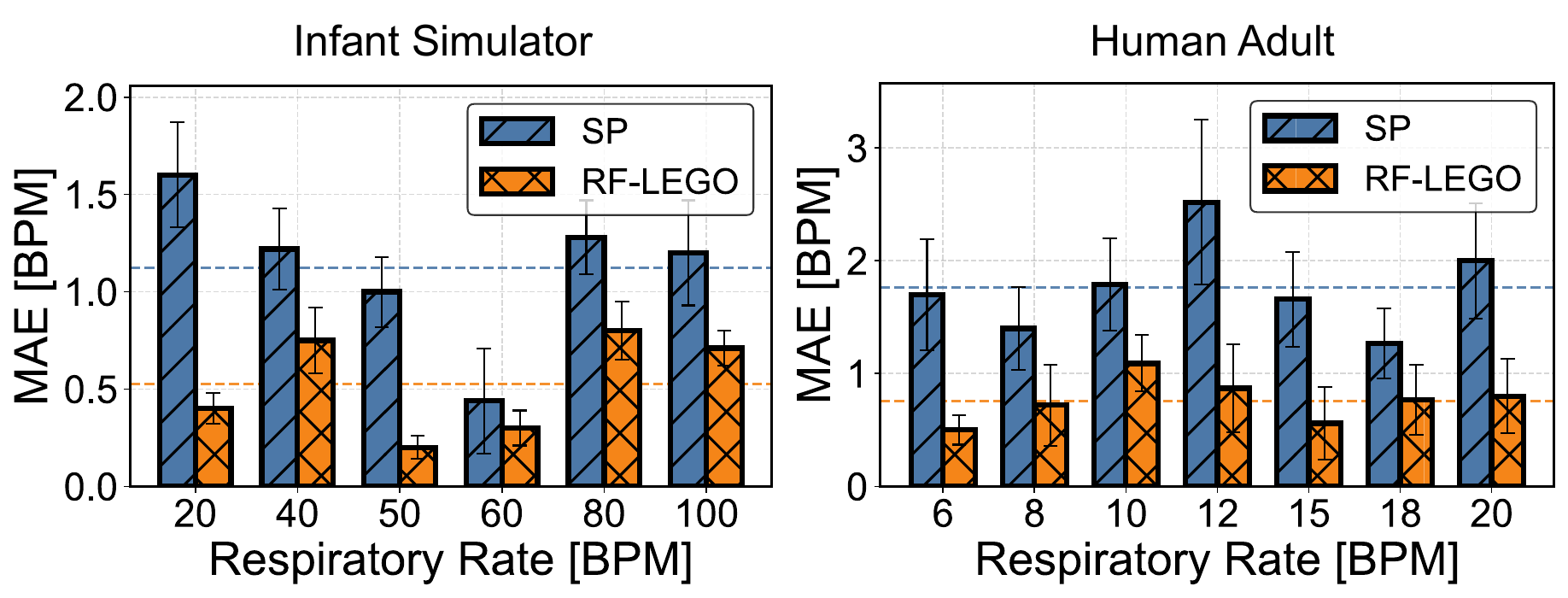}
    \caption{Performance for an infant simulator and the human adult.}
    \label{fig:case_study_breathing_results}
  \end{minipage}
  \begin{minipage}{0.32\textwidth}
  \centering
    \includegraphics[width=\linewidth]{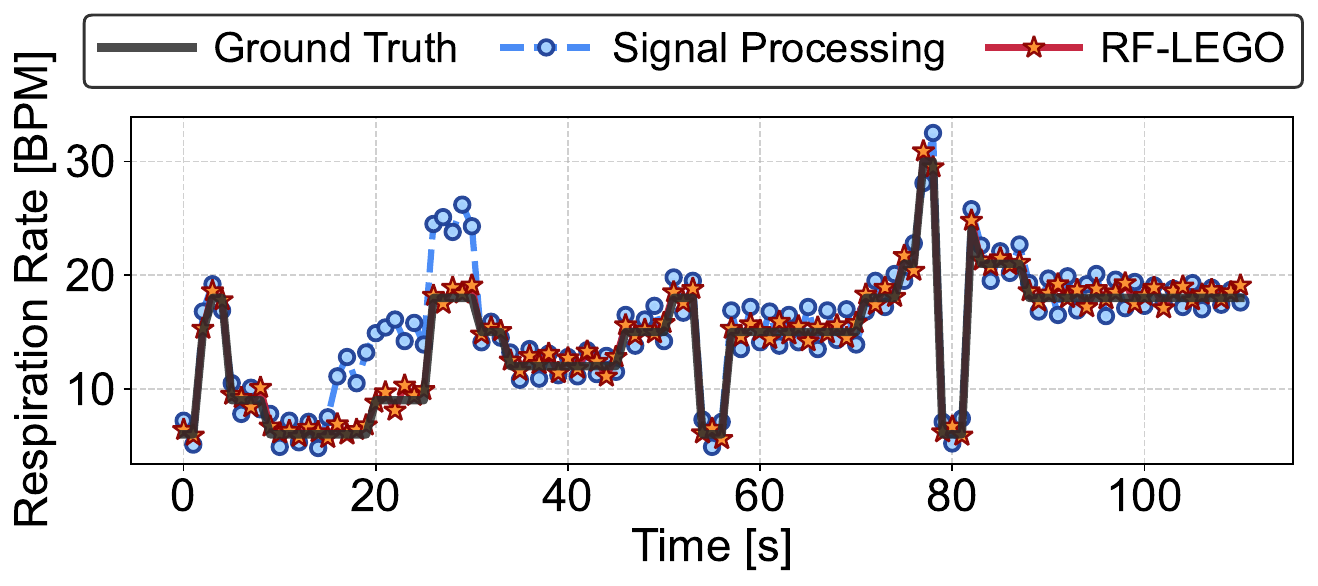}
    \caption{Natural breathing monitoring by a human adult.}
    \label{fig:case_study_breathing_sample}
  \end{minipage}
  \begin{minipage}{0.33\textwidth}
  \centering
    \includegraphics[width=\linewidth]{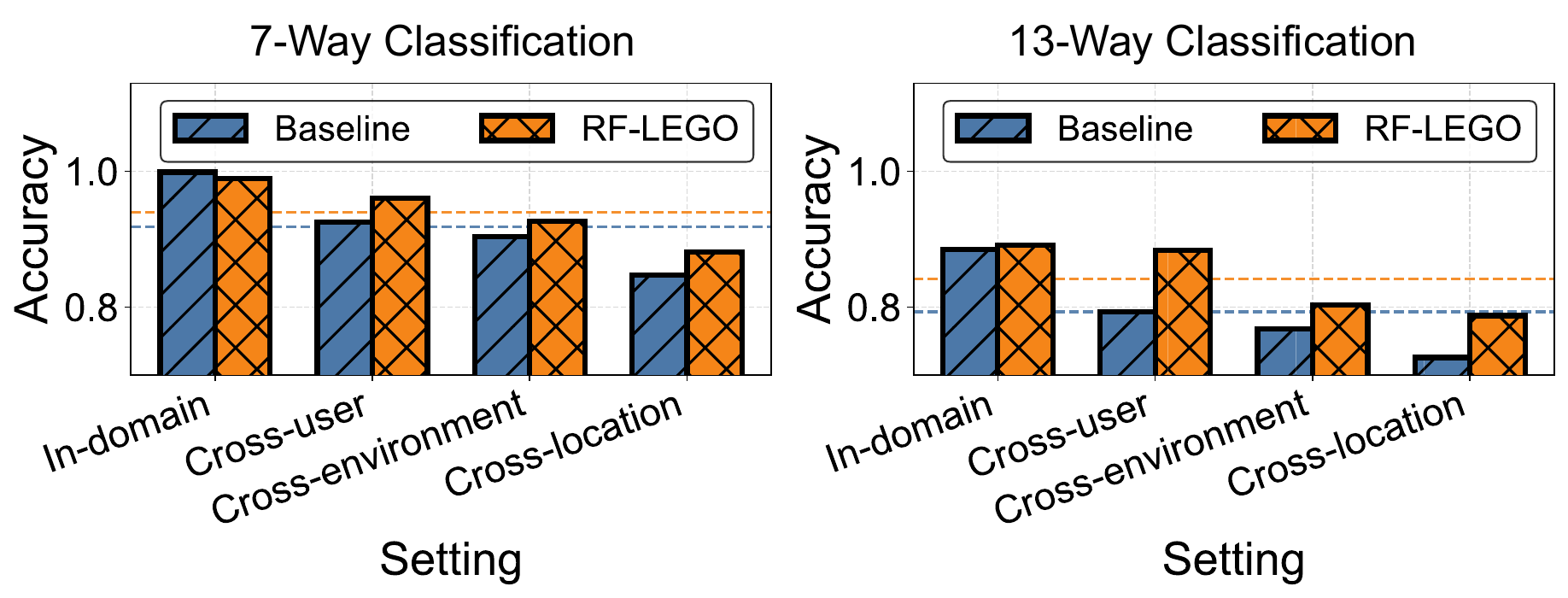}
    \caption{Accuracy for 7-way and 13-way classification across various settings.}
    \label{fig:case_study_har_results}
  \end{minipage}
\end{figure*}

\subsection{Trajectory Tracking}

\subsubsection{Experimental Design}
As shown in \fig \ref{fig:case_study_scenario}(a-b), we place the mmWave radar at the corner of a 4m$\times$4m room instrumented with an OptiTrack motion capture system \cite{optitrack} to obtain ground truth human trajectories. The dataset contains 65 designated trajectories spanning 13 categories, including straight, turn, circle, and 5 random trajectories. We compare two pipelines in \fig \ref{fig:case_study_pipeline}, keeping pre-processing, tracking, and hyperparameters identical (both processing and Kalman filter) to ensure a fair test. We use Absolute Trajectory Error (ATE) and Relative Trajectory Error (RTE) as the metric, where ATE indicates the RMSE of the predicted trajectory and the ground truth trajectory, and RTE measures the RMSE between both trajectories over a time interval. And we use a 1-second time window for evaluation.

\subsubsection{Evaluation}
The \sysname-integrated pipeline demonstrates superior performance in the trajectory tracking task. Quantitative results are presented in \fig \ref{fig:cdf_of_ate_rte}. Compared to the traditional SP baseline, the \sysname pipeline significantly reduces both error metrics. For instance, the median of ATE for \sysname is about 0.3 meters, a 40\% error reduction from the traditional baseline. Similar gains are observed for RTE, further confirming \sysname's high precision and stability throughout the tracking process. 
\fig \ref{fig:case_study_trajectory_sample} visualizes example tracking trajectories. As seen, the \sysname trajectories (red) closely align with the ground truth (gray). In contrast, the trajectories from the SP pipeline (blue) exhibit noticeable drift and greater deviation, especially when tracing complex letter shapes. This indicates that by providing a higher-quality, lower-noise point cloud, \sysname enhances the overall tracking robustness and precision.

\subsection{Vital Sign Monitoring}

\subsubsection{Experimental Design}
As shown in \fig \ref{fig:case_study_scenario}(c-e), we evaluate breathing in two setups: an infant simulator with programmable breathing rates and human adults instrumented with a respiratory belt for ground truth. The simulator is tested at \{20, 40, 50, 60, 80, 100\} BPM. Human trials span \{6, 8, 10, 12, 15, 18, 20\} BPM, plus spontaneous resting sessions with one and two participants. Each condition is recorded for 6 minutes. We compare the two pipelines in \fig \ref{fig:case_study_pipeline}, holding pre-processing, demodulation, and the rate estimator fixed to ensure a fair comparison. We measure the MAE of BPM as the evaluation metric.

\subsubsection{Evaluation}
\fig \ref{fig:case_study_breathing_cdf} presents the CDF of the MAE for breathing rate estimation, aggregating the results from both our single and multiple person experiments of scenarios d and e.
The result reveals that \sysname achieves an 80\%-tile MAE of less than 2 BPM, while the traditional baseline yields 5$\times$ higher error of 10 BPM. 
\fig \ref{fig:case_study_breathing_results} further details the performance on both an infant simulator and human subjects. Whether for the various breathing frequencies of the infant simulator (ranging from 20 to 100 BPM) or the human adult (ranging from 6 to 20 BPM), 
As seen, \sysname consistently outperforms the baseline in all cases, achieving a respective MAE reduction of 52.7\% and 56.8\% compared with the SP baseline for the infant simulator and human adults. 
\fig \ref{fig:case_study_breathing_sample} illustrates an example trace of an adult's natural breathing, where our \sysname-enhanced solution tracks the breathing rates accurately and continuously, while the baseline experiences drifts and artifacts.

\subsection{Human Activity Recognition}
\label{sec:human_activity_recognition}

\subsubsection{Experimental Design}
We further validate the end-to-end effectiveness and cascadability of \sysname on more complex sensing tasks, \ie, human activity recognition using the MCD-Gesture dataset \cite{10001175}, a cross-domain mmWave benchmark spanning 6 environments, 25 users, and 5 locations (\ie, user positions relative to the radar), with 6 predefined gestures and 7 additional labeled activities. We evaluate two tasks: 7-way classification (merging the 7 non-gesture activities into a single Non-gesture class, following the dataset protocol) and 13-way classification (recognizing all gestures and activities). 
The baseline adopts a standard signal processing front-end that forms radar cube representations for a neural classifier \cite{9894724}. In the \sysname-integrated pipeline, we replace only the corresponding upstream operators with \sysname modules, while keeping the classifier backbone identical across methods.
We consider four evaluation settings via train/test splits: in-domain (random 4:1 split), cross-user (15 users for training, 10 for testing), cross-environment (3 environments for training, 3 for testing), and cross-location (3 locations for training, 2 for testing).

\subsubsection{Evaluation}
\fig \ref{fig:case_study_har_results} summarizes the recognition accuracy across the four settings. \sysname consistently improves over the signal processing baseline in both tasks, indicating that the learned RF front-end produces features that are directly usable by a downstream classifier, proving \sysname's cascadability with neural blocks as well. Quantitatively, \sysname achieves a higher mean accuracy of $93.92\%$ compared with $91.88\%$ on 7-way classification, and $84.15\%$ compared with $79.31\%$ on the more challenging 13-way classification. The larger gain in 13-way recognition suggests that \sysname is beneficial when finer-grained class boundaries amplify the impact of front-end feature quality, while remaining compatible with the same neural back-end.

\section{Related Work}
\label{sec:related_work}

\noindent \textbf{Deep Wireless Sensing.}
Recent advances in deep wireless sensing (DWS) have seen the successful application of large-scale models to various tasks \cite{chen2021movi,zhao2023radio2text,zhao2025space,freegait-tmc,LiquImager,lai2024enabling,zheng2019zero,zhao2018rf,dodds2025non,zheng2021more,hou2024rfboost}. However, their black-box nature often overlooks the physical properties of RF signals, motivating a surge in signal processing-deep learning (SP-DL) co-design. Key strategies include directly embedding SP blocks into networks \cite{ding2020rf}, developing phase-aware complex-valued models \cite{yang2023slnet,chi2024rf}, adopting DSP-inspired mechanisms \cite{yao2019stfnets,li2021units,zhao2023cubelearn}, and designing interpretable, complex-valued transformers \cite{zhang2025unlocking}.

\noindent \textbf{Deep Unrolling.}
Deep unrolling, also known as algorithm unrolling, is a principled SP-DL co-design methodology for building interpretable neural networks by mapping traditional iterative algorithms into the hierarchical structure of a deep neural network \cite{monga2021algorithm}. The seminal work in this domain is the Learned Iterative Shrinkage-Thresholding Algorithm (LISTA) \cite{gregor2010learning}, which successfully unrolled the ISTA algorithm for sparse coding into a feed-forward network. This foundational concept has since been successfully applied across diverse domains, from image processing tasks like restoration and deblurring \cite{chen2016trainable,li2020efficient,solomon2019deep} to RF sensing with models such as KalmanNet \cite{revach2022kalmannet} and DA-MUSIC \cite{merkofer2023music}. Its applicability further extends to solving differential equations \cite{chen2018neural,long2018pde}, demonstrating its versatility.
\sysname pioneers modular SP-DL co-design via deep unrolling for trustworthy RF sensing.

\section{Discussion}
\label{sec:discussion}

While our work demonstrates that \sysname successfully bridges the gap between classical signal processing and deep learning, \sysname also reveals a limitation that merits further investigation: information alienation. Although the unrolled modules perform exceptionally well on downstream tasks, their learned internal representations may diverge significantly from the clean, canonical outputs with clear physical definitions produced by traditional algorithms. This indicates that our interpretability claim is primarily structure-aligned: we preserve well-defined input/output contracts and semantically meaningful intermediate outputs, but we do not guarantee full fidelity or interpretability of all internal learned states within each block.
A key future direction is balancing physical fidelity with task optimization to achieve fully interpretable, high-performance sensing.

\section{Conclusions}
\label{sec:conclusion}

In this paper, we address a fundamental dilemma in RF sensing that forces a choice between rigid, interpretable signal processing pipelines and adaptable but opaque deep learning models. We introduce \sysname, a novel signal processing-deep learning co-design approach that resolves this trade-off by unrolling cornerstone algorithms, including frequency transform, beamforming, and detection, into trainable, physically-grounded neural operators. 
Our extensive evaluations demonstrate that these modules not only outperform their traditional counterparts in robustness and accuracy but also compose seamlessly into end-to-end pipelines, significantly enhancing downstream applications. By providing a library of interpretable, reusable \emph{LEGO bricks}, \sysname opens up a new paradigm of designing flexible, robust, and trustworthy solutions for RF sensing.

\begin{acks}
    This work was supported by NSFC under grant No. 62222216, Hong Kong RGC ECS No. 27204522, GRF No. 17212224 and No. 17211725, and CRF No. C5002-23Y. We thank the anonymous Reviewers and Shepherd for their insightful feedback.
\end{acks}

\bibliographystyle{ACM-Reference-Format}
\bibliography{refs}

\end{document}